\documentclass[aps,pre,twocolumn,showpacs,superscriptaddress]{revtex4-1}
\usepackage{float}
\usepackage{lineno}
\usepackage{comment}
\usepackage{lipsum}
\usepackage{mathrsfs}
\usepackage{graphicx}
\usepackage{wrapfig}
\usepackage{bm}
\usepackage{amsmath,amssymb}
\usepackage{mwe}   
\usepackage{subfig}
\usepackage{color}
\usepackage[table]{xcolor}
\definecolor{truegreen}{RGB}{8, 148, 4}
\definecolor{irishgreen}{rgb}{0, 0.565, 0.165}
\usepackage[colorlinks = true,linkcolor=black, citecolor=irishgreen,urlcolor =irishgreen]{hyperref}
\usepackage{stackengine}
\usepackage[font=normalsize] {caption}
\usepackage[section]{placeins}
\usepackage[english]{babel} 

\makeatletter
\newcommand*{\rom}[1]{\expandafter\@slowromancap\romannumeral #1@}
\makeatother

\renewcommand{\v}{{\bm v}}
\renewcommand{\u}{{\bm u}}

\newcommand{\vi}{{\bm  v}_i}
\newcommand{\ri}{{\bm r}_i}
\newcommand{\av}[1]{\left\langle #1 \right\rangle}

\graphicspath{{figures/}}

\begin{document}

\title{Hydrodynamics of Immiscible Binary Fluids with Viscosity Contrast:\\ A multiparticle collision dynamics approach}
\author{Zihan Tan}\email{z.tan@fz-juelich.de}
\affiliation{Biomacromolecular Systems and Processes, Institute of Biological Information Processing, Forschungszentrum J\"ulich, 52428 J\"ulich, Germany}
\author{Vania Calandrini}
\affiliation{Computational Biomedicine, Institute for Advanced Simulation, Forschungszentrum J\"ulich, 52428 J\"ulich, Germany}
\author{Jan K. G. Dhont}
\affiliation{Biomacromolecular Systems and Processes, Institute of Biological Information Processing, Forschungszentrum J\"ulich, 52428 J\"ulich, Germany}
\affiliation{Heinrich-Heine Universit\"at D\"usseldorf, Department of Physics,
D-40225 D\"usseldorf, Germany}
\author{Gerhard N\"agele}
\affiliation{Biomacromolecular Systems and Processes, Institute of Biological Information Processing, Forschungszentrum J\"ulich, 52428 J\"ulich, Germany}
\affiliation{Heinrich-Heine Universit\"at D\"usseldorf, Department of Physics,
D-40225 D\"usseldorf, Germany}
\author{Roland G. Winkler}
\affiliation{Theoretical Physics of Living Matter, Institute for Advanced Simulation, Forschungszentrum J\"ulich, 52428 J\"ulich, Germany}


\begin{abstract}

We present a multiparticle collision dynamics (MPC) implementation of layered immiscible fluids $A$ and $B$ of different shear viscosities separated by planar interfaces. The simulated flow profile for imposed steady shear motion and the time-dependent shear stress functions are in excellent agreement with our continuum hydrodynamics results for the composite fluid. The wave-vector dependent transverse velocity auto-correlation functions (TVAF) in the bulk-fluid regions of the layers decay exponentially, and agree with those of single-phase isotropic MPC fluids. In addition, we determine the hydrodynamic mobilities of an embedded colloidal sphere moving steadily parallel or transverse to a fluid-fluid interface, as functions of the distance from the interface. The obtained mobilities are in good agreement with hydrodynamic force multipoles calculations, for a no-slip sphere moving under creeping flow conditions near a clean, ideally flat interface. The proposed MPC fluid-layer model can be straightforwardly implemented, and it is computationally very efficient. Yet, owing to the spatial discretization inherent to the MPC method, the model can not reproduce all hydrodynamic features of an ideally flat interface between immiscible fluids.
\end{abstract}

\maketitle

\section{Introduction}
Multi-phase fluid flows occur ubiquitously in nature and engineering processes. Examples constitute oil-water flows, fluids with air bubbles, emulsions, dairy products, biological fluids, processing of paints, coating, and printing. Owing to their complexity,  the theoretical description and efficient modeling of binary fluids pose major challenges, which stimulated a wealth of endeavors to model binary fluids using mesoscale simulations. In the lattice-Boltzmann method (LBM), the implementation of multi-phase flows and phase separation encompasses several variants: the color gradient model~\cite{rothman1988,Gunstensen1991,Yu2019}, the pseudo-potential model~\cite{Shan1993,Shan1994}, the free-energy functional model~\cite{Orlandini1995,swift1996}, and the mean-field model~\cite{He1998}, or combinations thereof~\cite{Lombard2020}. Dissipative particle dynamics (DPD) simulations, which explicitly account of conservative pair interactions between fluid particles, allow to realize multi-phase fluids via assigning distinct interactions between the particles~\cite{Coveney1997}. Furthermore, the multiparticle collision dynamics (MPC) method, a particle-based hydrodynamic simulation approach which captures hydrodynamic interactions and thermal fluctuations~\cite{kap99,MPCD,kapral_review,huang12,huan:13,yang15amc,theers16friction}, has been proven valuable and efficient for mesoscale simulations, and has been applied in a broad range of studies of biological  and active polymers~\cite{liebetreu2018,gomez2020,kiku:02,jcp05,chel:12,ryde:06,fran:08,huan:10,ripoll:06}, colloids \cite{sing:12,sing:11,padding2004,ripoll2008,tan01}, proteins~\cite{bucciarelli2016dramatic,das2018clustering}, vesicles and blood cells \cite{nogu:04,nogu:05}, microswimmers~\cite{hu15,wagner2017hydrodynamic,qi2020,tao:10,zoet:12,babu:12,Elgeti19032013,reig:12,thee:14,yang14gear,mous:20}, and microfluidics~\cite{16ratchet,tan02}. To date, various MPC implementations of binary fluid mixtures have been proposed, and their phase behavior has been studied~\cite{hashimoto2000,lnoue2004,Ihle2006,Tuezel2007,tuezel2010,echeverria2017,hill:16,eise:18}. 
Depending on the applied interaction rule between the different fluid components, the viscosity values of the (two) fluids are equal or individually controlled \cite{Tuezel2007,hill:16,mesl:09,goe07}.

Most of the above mentioned simulation methods are aimed to account for both the hydrodynamics and thermodynamics. The large computational costs, which are often necessary to suitably account for the thermodynamics involved in studying phase separation of multi-phase fluids, are dispensable when the dynamics of embedded objects such as proteins, polymers, or living organisms are considered. In fact, a plethora of physical phenomena related to immiscible binary fluids take place under conditions where phase separation is absent or is of no interest, and simulation methods accounting for the hydrodynamics alone suffice here. The MPC approach is very well suited to efficiently simulate hydrodynamic flow properties in the presence of thermal fluctuations (fluctuating hydrodynamics) \cite{huang12}. In particular, MPC  allows to tune the viscosity of fluids through the specification of the frequency of MPC collisions, and, hence,  to control the viscous properties of immiscible fluids. 

In this work, we present a model for planar layers of two immiscible binary fluids $A$ and $B$ using the MPC  approach. The fluids, separated by a flat interface,  are of distinct shear viscosity, $\eta_A$ and $\eta_B$, whose values are tuned by the corresponding MPC collision frequency. While omitting the thermodynamic and kinetic processes of phase separation, it allows for fluid particle exchange across the interface, associated with a change of the local (collisional) interactions in the arriving fluid layer. No explicit interactions between fluid particles at the interfacial zone are required, although a more sophisticated modeling of the interface properties is possible for future assessment. Shear flow profiles and  the shear stress under starting flow conditions are calculated, and the latter is compared with a provided analytical solution of the linearized Navier-Stokes equation of the same composite fluid. 
Moreover, transverse hydrodynamic velocity correlation functions (TVAFs) are determined for the different layers. In addition, the hydrodynamic mobility/friction properties of a colloidal sphere inside a fluid layer, which moves steadily parallel or perpendicular to an interface, are calculated. The approach recovers the correct flow profiles, fluctuating hydrodynamic properties, and thermal fluctuations of the individual fluid layers.  The invoked simplifications in the present MPC treatment lead to a higher computational efficiency compared to other mesoscale  techniques and MPC implementations considered so far \cite{Ihle2006,Tuezel2007,tuezel2010,echeverria2017,hill:16,eise:18}, which is a significant advantage when simulating large-scale systems.

The present simulations constitute a first important step in studying the dynamics, e.g., of monolayers of thermal particles moving near a planar fluid-fluid interface, with full account of the time-resolved (retarded) hydrodynamic interactions of the particles with the interface and among each other. These so-called quasi-two-dimensional systems have been intensely studied recently, since they reveal peculiar dynamic features such as the anomalous hydrodynamic enhancement of lateral collective diffusion \cite{naegele2001jcp,naegele2002,panzuela2017collective,panzuela:2018}, and the influence of the interface on the motion  of nearby Brownian particle, as reflected in the non-isotropic, hydrodynamic long-time tails of particle velocity correlations \cite{feld:2005,huan:2015}. Interestingly enough, the motion pattern of microswimmers is also strongly affected by their hydrodynamic interaction with a nearby (fluid) interface \cite{lope:2014,hu:15}.

Our two-fluids MPC model is also a first step toward mesoscopic simulations of the diffusion and phase behavior of  assemblies of interacting proteins attached to or embedded inside a membrane. It should be recognized here that the biophysical properties of the membrane, both in physiological and in vitro conditions, influence the structure and function of many membrane-associated proteins~\cite{janm:2006,kapo:2019,corr:2013,rome:2019,rebe:2017,oate:2011,mann:2016}. Diffusion properties of single membrane receptor proteins and their orientation-dependent interaction potentials (which can be partially due to local membrane deformations) as obtained from force-field based molecular dynamics (MD) simulations where the lipids and the atomistic structure of the receptor are explicitly accounted for, can be used as input to tune mesoscopic MPC simulations.

The present paper is organized as follows. Section~\ref{sec:model} gives the essentials of the single-phase MPC algorithm, outlines its extension to immiscible multi-phase fluids, describes the coupling rules of a colloid with the MPC fluid, and defines the simulation parameters. The two-fluids MPC model is validated in the three subsequent sections. In section~\ref{sec:shear}, the stationary shear profile of the planar three-layers system and the time-dependent  shear stress functions under starting flow conditions are simulated, and compared with our analytic continuum hydrodynamics results. In section~\ref{sec:hydro_corr}, MPC simulated transverse velocity correlation functions (TVCFs) in the bulk regions of the two fluids, and in a region including the interface between them, are contrasted with predictions from the linearized fluctuating Landau-Lifshitz Navier-Stokes equation. In Section~\ref{sec:mobility}, the simulated hydrodynamic mobilities of a colloidal sphere moving steadily inside the middle layer of a three-layers fluid system are compared to previous numerical results based on the Stokes equation of low-Reynolds number hydrodynamics. In section~~\ref{sec:suc} we summarize and conclude our findings, and provide a perspective on future work. The Appendix presents our continuum hydrodynamics results for the time-dependent velocity profiles and for the stress functions of the composite three-layers system under starting shear flow conditions.

\section{Model} \label{sec:model}
\subsection{Multiparticle collision dynamics (MPC) fluid}

A single-phase MPC fluid consists of $N$ point particles each of mass $m$, typically enclosed in a cubic simulation box of length $L$ with periodic boundary conditions. The  dynamics of the fluid particles proceeds through discrete streaming and collision steps \cite{kap99,kap00,kapral_review,MPCD}. During a streaming step, the particles move ballistically for a time span $h$, referred to as collision time. Hence, the position $\ri$ of a fluid particle $i$, with $i\in\{1,\ldots,N\}$, is updated according to 
\begin{align} 
	\ri(t+h)=&\ri(t)+h \v_i(t)\,,
	\label{eq1}
\end{align}
where $\v_i$ is the particle velocity. In the subsequent collision step, accounting for the interactions between fluid particles, the MPC particles are sorted into cubic cells of size $a$ defining the local interaction environment (collision cells). In the stochastic rotation dynamics (SRD) variant of MPC, MPC-SRD, the relative  particle velocities $\tilde{\v}_i=\vi-\v_{cm}$, with respect to the center-of-mass velocity $\v_{cm}$ of a particular collision cell, are rotated around a randomly orientated axis by a fixed angle $\alpha$~\cite{MPCD,kapral_review,ihl01}. In three dimensions, the velocity of a particle $i$ after a collision is thus given by
\begin{align}
	\vi(t+h)=& \v_{cm}(t)+ \tilde{\v}_{i,\perp}(t)\cos(\alpha)\\ &  +  [\tilde{\v}_{i,\perp}(t) \times \Re] \sin(\alpha)+ \tilde{\v}_{i,\parallel}(t)\,,
	\label{eq2}
\end{align}
where  $\Re$ is a  unit vector along the selected rotation axis, and $\tilde{\v}_{i,\perp}$ and $\tilde{\v}_{i,\parallel}(t)$ are the parallel and perpendicular components of the velocity $\tilde{\v}_i$ with respect to the vector $\Re$, respectively. The orientation of $\Re$ is chosen independently for each collision cell and time step.  
The MPC-SRD algorithm conserves particle number, energy, and linear momentum. However, angular momentum is not conserved~\cite{theers16friction}, an aspect irrelevant for the current study~\cite{Winkler2009,huang12,winkler2011backtracking}. The employed discretization into collision cells breaks Galilean invariance, which is re-established  by a random shift of the collision cell lattice at every collision step \cite{ihl03}. To maintain a constant temperature, a (simple) cell-level  scaling scheme of the relative velocities $\tilde{\v}_i$  is employed \cite{huang2015}. Since mass, momentum and energy are conserved locally, the correct fluctuating hydrodynamic equations for an isothermal compressible fluid are obtained in the continuum limit. 

The shear viscosity $\eta =\eta^k + \eta^c$ of a homogeneous MPC fluid can approximately  be calculated analytically and   comprises contributions from streaming, $\eta^k$, and collisions, $\eta^c$, \cite{tuezel2006,nog07b,Winkler2009,huang2015,theers2015} where  
\begin{align}
	\eta^k = & \ \frac{\av{N_c} k_B T h}{2 a^3} \left[ \frac{5 \av{N_c}}{(\av{N_c} -1)(2 - \cos \alpha -\cos 2\alpha)}\right] , \\
	\eta^c = & \ \frac{\av{N_c}m}{18 a h} \left[ 1 - \cos \alpha\right] \left(1 - \frac{1}{\av{N_c}} \right) .
\end{align}
Here, $\langle N_c \rangle$ is the average number of MPC particles per collision cell, $T$ the temperature, and $k_B$ the Boltzmann constant.  
Note that the kinetic contribution, $\eta^k$, is only approximately valid, because its derivation is based on the molecular chaos assumption \cite{huang2015}. However, for small collision time steps, $\eta$ is dominated by $\eta^c$.

The viscosity of a MPC fluid  can be adjusted in several ways: by changing the mass of a particle, the average particle density in a collision cell, the rotation angle $\alpha$, and the collision time step $h$.  An implementation of two fluids with different masses has been realized  \cite{mesl:09,goe07}. Since the particles are distinctly different, suitable boundary conditions between the two fluids have to be applied. Similarly challenging are simulations  of coexisting fluids with a MPC particle density difference. Variations  of the collision angle yield a rather limited range of viscosity differences \cite{pre05}.  The most suitable strategy to simulate  immiscible fluids seems to be a change of the collision time step in the different fluid regimes. In fact, the particles themselves are identical in the various fluids, they only experience more (or less) frequent collisions. This drastically simplifies the numerical implementation and enhances the performance.

\subsection{Immiscible binary fluid system}

\begin{figure}[t]
	\centering
	\includegraphics*[width=0.40\textwidth]{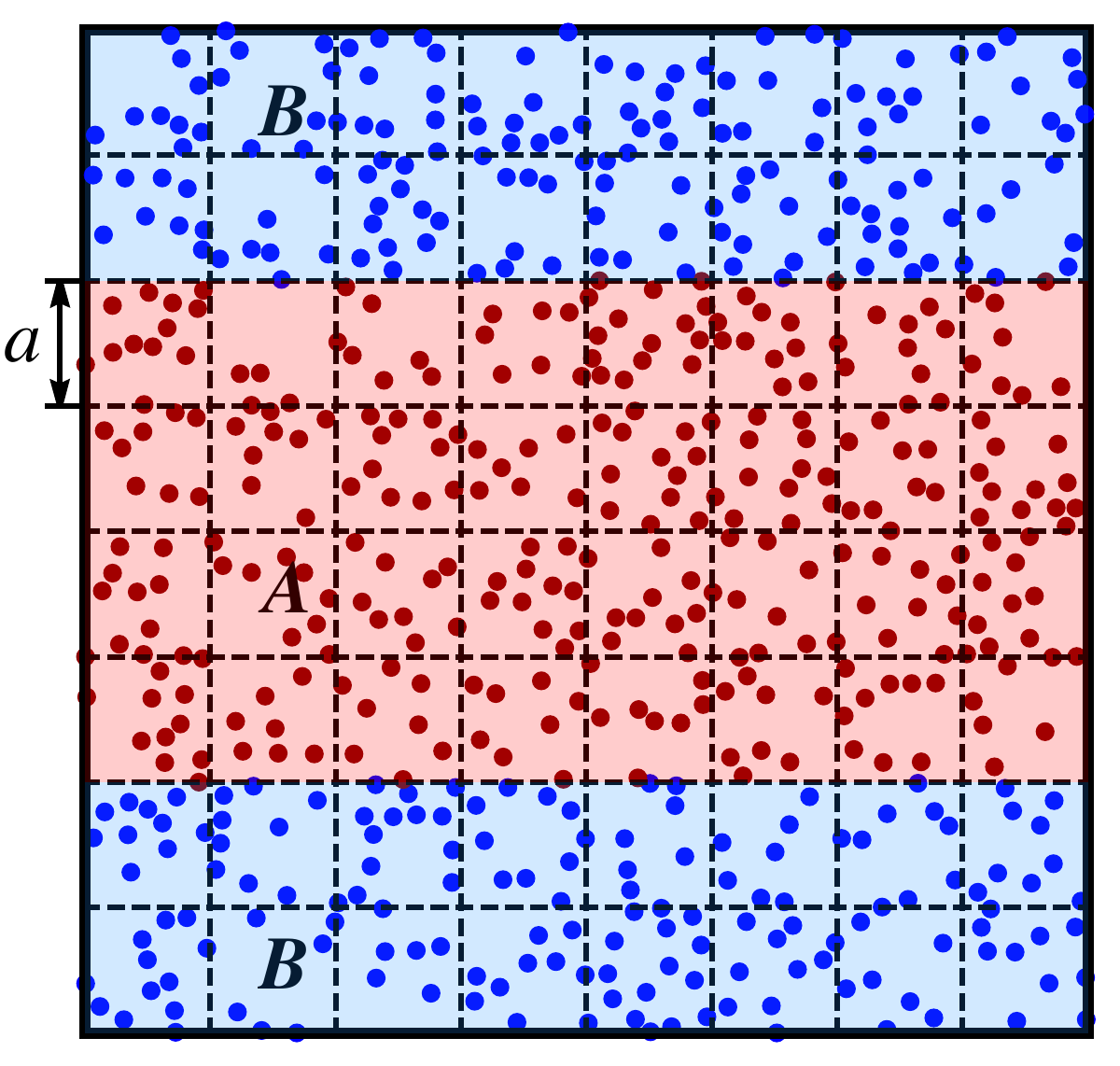}
	\caption{\label{fig:Sc}Schematics of a periodic three-layer MPC system of two immiscible fluids $A$ (red) and $B$ (blue) separated by planar intefaces.  MPC collisions are performed independently in cubic collision cells, delineated by the dashed lines, using  collision times $h_A$ and $h_B$ according to the fluid type. The width, $a$, of a cubic MPC cell is indicated.}
\end{figure}

The extension from a single to a layered two-fluid system of phases $A$ and $B$ with distinct shear viscosities $\eta_A$ and $\eta_B$, which are separated by two flat interfaces, is rather straightforward and illustrated in Fig.~\ref{fig:Sc}. As discussed above, the viscosity of a single-phase MPC fluid is sensitive to the collision time $h$~\cite{MPCD,Winkler2009}, and we describe the fluid layers $A$ and $B$ of  different viscosities by using accordingly different collision time steps $h_A$ and $h_B$. Without loss of generality, we take $h_A < h_B$ in the following, implying the viscosity of fluid $A$ to be larger than that of fluid $B$. Furthermore, $A$ and $B$ particles are assumed to be of equal mass $m$, and the mean mass densities of both fluid phases are taken to be the same. Hence, the MPC particles in the two regimes are identical. For notational convenience, we will refer to  particles $A$ and $B$ and fluids $A$ and $B$ corresponding to MPC particles in the two different domains.

As for a single-phase fluid, the MPC fluid particles move ballistically, and undergo independent collisions as follows:
\begin{itemize}
	\item[(1)] Streaming: $A$ and $B$ fluid particles move according to Eq.~\eqref{eq1} with collision time step $h_A$. This ensures an identical  ``continuous'' dynamics of both particle types. 
	\item[(2)] Sorting: After streaming with time $h_A$, the particles are sorted in (shifted) collision cells.
	\item[(3)] Rotation: Particles in the collision lattice of the $A$ domain undergo rotations according to Eq.~\eqref{eq2}, after the time interval $h_A$. Particles in the collision lattice  of the $B$ domain are rotated only after the time interval $h_B> h_A$, i.e., after $h_B/h_A$ additional streaming steps of time length $h_A$. 
\end{itemize}  
The subdivision of the streaming of particles $B$ does not affect the properties of the bulk part of fluid $B$, since the particles  move ballistically. However, it is important for the particles close to the planar $A$-$B$ interface, owing to the random shift of the collision lattice  normal to the interface, and since all particles in the (shifted) lattice of the $A$ fluid domain undergo rotations after the time interval $h_A$. The  equivalent streaming motion of particles in the two domains ensures an, on average, homogeneous distribution of fluid particles across the system, and a homogeneous density across the interface. However, the random shift of the collision cell lattice normal to an $A$-$B$ interface broadens the interface at least by the size $a$ of a collision cell.

\subsection{Colloid dynamics and fluid coupling} \label{sec:coupling}

The translational and rotational motions of a neutrally buoyant no-slip hard-sphere colloid of radius $R$ and mass $M$ embedded in the MPC fluid is governed by elastic collisions with the MPC particles, which we account for in a  coarse-grained manner \cite{theers16friction}. During the streaming step, just as for the solvent particles, a colloidal sphere moves ballistically with center-of-mass velocity  $\bm V_c(t)$. Its center-of-mass position vector, $\bm R_c(t)$, changes according to 
\begin{align}
	\bm{R}_c(t+h)&=\bm{R}_c(t)+ h \bm{V}_c(t) .
	\label{eq:stream_Vc}
\end{align}
MPC particles $i$ which (virtually) penetrate the colloid are moved backwards in time by the time interval $(h-h_i)$, where $h_i<h$ follows from the condition $|\bm r_i(t)-\bm R_c(t)+h_i(\v_i(t)-\bm V_c(t))|^2=R^2$. These MPC particles collide then with a virtual colloid at the center position $\bm R_c(t) + h_i \bm V_c(t)$, transfer the momentum $\bm p_i$ elastically to the colloid, and subsequently move with new velocity $\bm v_i'=\bm v_i(t+h_i)$ for the time interval $(h-h_i)$.
The linear and angular velocities of the MPC particles and of the colloid  before and after collision are related by
\begin{align} \label{eq:collide_v} 
	\v_{i}' & =\v_{i}(t)- \bm p_i/m , \\  \label{eq:collide_Vc}
	\bm{V}_c (t+h) & =  \bm{V}_c(t)+ \sum^{}_{i=1}\bm{p}_i/M , \\
	\bm{\Omega}_c(t+h) &  = \bm{\Omega}_c(t)+R \sum^{}_{i=1}(\bm n_i\times \bm p_i) / I\,.
	\label{eq:collide_ang_v}
\end{align}
The sum extends over all fluid particles colliding with the colloid during the time interval $h$. Here, $\bm \Omega_c$ is the angular velocity of the embedded colloid, $\bm n_i = (\bm r_i-\bm R_c)/\left|\bm r_i-\bm R_c \right|$ is the unit vector pointing from the colloid center to the position of fluid particle $i$, and $I=\chi M R^2$ with $\chi=2/5$ is the moment of inertia of the spherical colloid. 

To realize the hydrodynamic no-slip boundary conditions at the colloid surface, 
we use the  bounce-back rule for the MPC fluid particles, which yields~\cite{padding2005stick,zottl2014,theers2016modeling,theers16friction,theers2018clustering}
\begin{align}
	\bm{p}_i=2\mu_\text{red}\breve{\v}_{i,n}+2\mu_\text{red} \frac{M\chi}{\mu_\text{red} +M\chi}\breve{\v}_{i,t}\,,
	\label{eq:pi}
\end{align}
with the relative velocity, $\breve{\v}_{i}$, of a colliding MPC fluid particle $i$ with respect to the according colloid surface point given by
\begin{align}
	\breve{\v}_i =\v_i-\left[\bm{V}_c+R\;\!\bm{\Omega}_c\times \bm{n}_i\right]\,.
	\label{eq:vtilde}
\end{align}
Here, $\mu_\text{red}=mM/(m+M)$ is the reduced mass, and $\breve{\v}_{i,n}$ and $\breve{\v}_{i,t}$ are the normal and tangential relative velocity parts, respectively, with respect to the colloid surface.

In the MPC collision step, phantom (p)  particles are added inside the colloid to enforce the no-slip hydrodynamic boundary condition, which, in addition,  act as a thermal bath \cite{Lamura01}. Theses particles are uniformly distributed inside the colloid according to the average MPC-fluid particle density, and their velocities relative to the colloidal translational and rotational velocities are taken from a central Maxwellian distribution function. This yields the updated colloid translational and angular velocities after a collision step
\begin{align}
	&\bm V_c(t+h) \rightarrow \bm V_c(t+h) + \sum_i \Delta\bm p_i^{p}/M , \\
	&\bm \Omega_c (t+h) \rightarrow \bm \Omega_c (t+h) + R \sum_i \left( \bm r_i^p - \bm R_c \right) \times \Delta\bm p_i^p/I \, , 
\end{align}
respectively. 
Here, $\Delta\bm p_i^p$ denotes the change in the linear momentum of phantom particle $i$ at position $\bm r_i^p$ due to SRD, and $\bm V_c(t+h)$ and $\bm \Omega(t+h)$ are the velocities in Eqs. \eqref{eq:collide_Vc} and  \eqref{eq:collide_ang_v}, respectively.

To further speed up the simulations, we use a common value $h_i=h/2$ for all MPC particles rather than considering  the individual elastic collision events at the exact times $t + h_i$ of each fluid particle-colloid collision. This simplifying step was shown to be as accurate as when the exact $h_i$s are used, especially for small collision time steps~\cite{hecht2005,padding2005stick,theers2016modeling}.

\subsection{Simulation parameters} \label{sec:sim_para}

In what follows, lengths are measured in units of $a$, mass in units of $m$, and energy in units of the thermal energy $k_{B}T$. We use therefore the units
\begin{align}
	t_0=\sqrt{ma^2/(k_{B}T)} , \hspace*{0.5mm}  
	v_0 = v_{th}= \sqrt{k_BT/m} , \nonumber \\
	\hspace{-5mm}
	\eta_0=\sqrt{m k_BT}/a^2
	\label{eq:units}
\end{align}
for time, $t_0$, velocity, $v_0$, and viscosity, $\eta_0$, respectively. Note that $t_0$ is equal to the ratio of cell size $a$ and thermal velocity $v_{th}=\sqrt{k_BT/m}$. In these units, the  sound velocity in both fluids is equal to one.
The average number of particles per collision cell is selected as $\av{N_c}=10$, implying equal mean number and mass densities of the $A$ and $B$ fluids, and the rotation angle is set to $\alpha=130^\circ$. The collision time steps are taken as $h_A=h_B/5=0.02\times t_0$.  With the  mass density $\rho= \langle N_c\rangle\;\!m/a^3$, the corresponding kinematic viscosities are $\nu_A/\nu_0=4.12$ and $\nu_B/\nu_0=0.87$, where $\nu_0 = \eta_0a^3/m$. This yields the kinematic viscosity ratio $\mu^2 = \nu_A/\nu_B= 4.74$. The related dimensionless Schmidt numbers are $Sc_A =\nu_A/D_A=400$ and $Sc_B =\nu_B/D_B=17$, expressing that the viscous diffusion of (transversal) momentum in the fluid is distinctly faster than diffusive mass transport, with the latter characterized by the mass diffusion coefficients $D_A=(h_A/h_B)D_B$ of fluid $A$ and $B$ particles, respectively. Simulations are performed using periodic boundary conditions, applied in sections~\ref{sec:shear} and \ref{sec:hydro_corr} to a cubic simulation box of length $L/a = 39$ and $80$, respectively, and in section~\ref{sec:mobility} to a rectangular box of lengths $2\;\!L_x/a=2\;\!L_y/a=L_z/a =80$. The latter embeds a colloidal sphere of radius $R=2.5a$.

\section{Shear simulations} \label{sec:shear}
\begin{figure}[h]
	\includegraphics[width=0.45\textwidth,trim={0 0 0 0},clip]{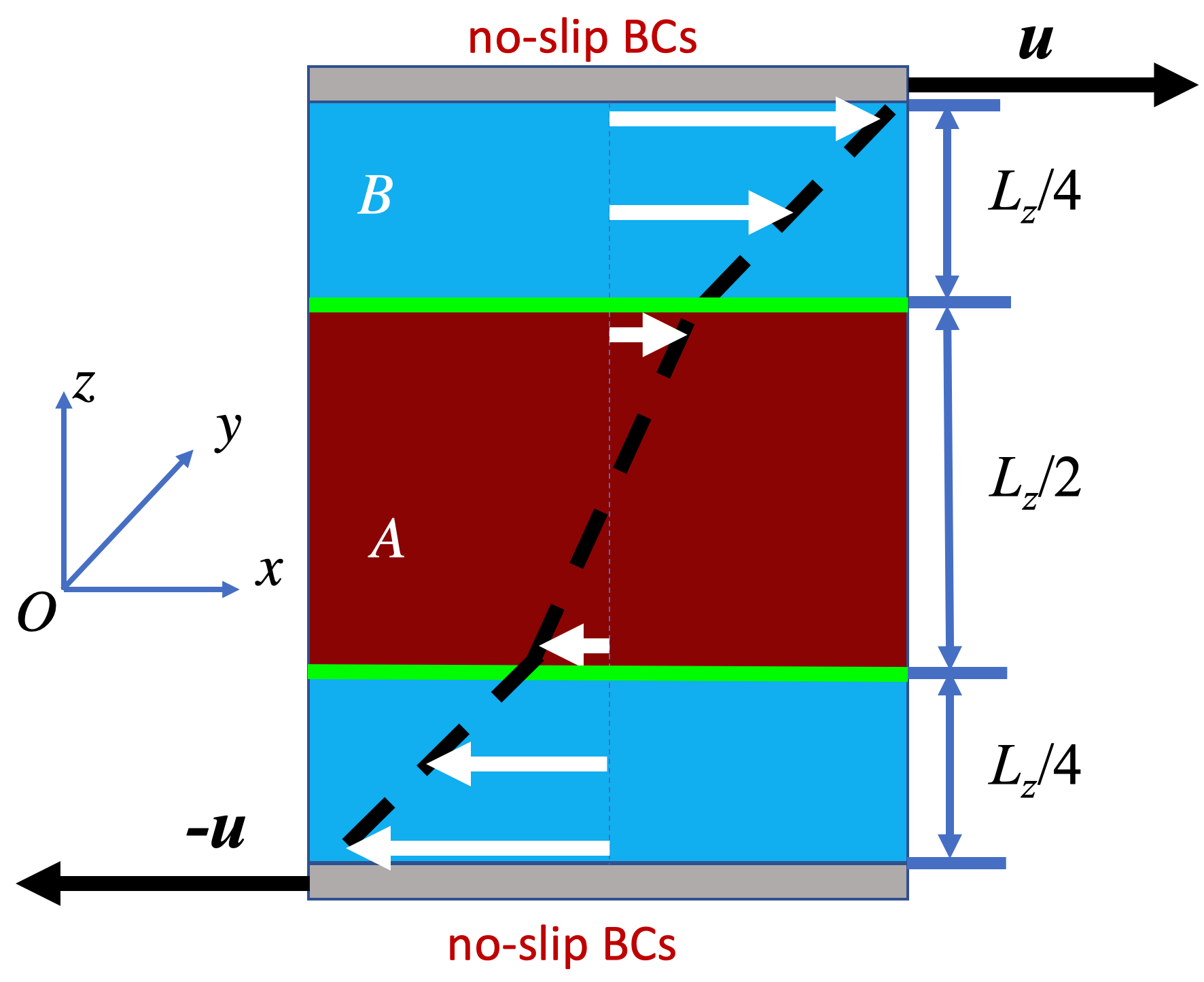}
	\caption{\label{fig:shear_sch} Schematics of a layered $B$-$A$-$B$ fluid system steadily sheared by two parallel no-slip walls  moving in  opposite direction with velocities $\pm\u=(\pm u,0,0)$. A piece-wise linear stationary velocity profile is obtained from hydrodynamics under laminar flow conditions.}
\end{figure}

As a first example used for scrutinizing the hydrodynamic behavior of our two-fluids MPC approach, we consider a standard stationary shear flow setup as sketched in Fig.~\ref{fig:shear_sch}. The three planar layers of two immiscible fluids $A$ and $B$ are sheared by two walls oriented parallel to the $xy$-plane, which move oppositely along the $x$-direction with constant velocities  $\pm u=0.0975\;\!v_{th}$. The lower wall is located at $z=0$, and the upper one at  $z=L_z = L =  39a$. No-slip boundary conditions (BCs) at the walls are implemented using the bounce-back rule and phantom particles inside the walls \cite{Winkler2009}. The three $B$-$A$-$B$ fluid layers are separated by planar fluid interfaces located at $z=L_z/4$ and $3L_z/4$, respectively. 

The stationary shear velocity ${\bm v}^{st}(z)=v^{st}_x(z)\;\!{\bm e}_x$, obtained from the stationary Navier-Stokes equation \cite{dhont1996introduction}, is piecewise linear and unidirectional along the $x$-direction. The flow is uniquely determined by the wall-fluid stick boundary conditions, and the continuity of flow velocity and shear stress across the two clean planar interfaces whose thickness is assumed to be zero. Explicitly,
\begin{align} \label{eq:bcs}
	v_x^{B,st}(L_z) & =-v_x^{B,st}(0)=u, \\  v_x^{A,st}(L_z/4) & = v_x^{B,st}(L_z/4)=u_{s}^-, \\ v_x^{A,st}(3L_z/4) & =v_x^{B,st}(3L_z/4)=u_{s}^+,
\end{align}
\begin{figure}[H]
	\centering
	\subfloat{		
		\begin{picture}(100,180)
			\put(-78,-15){\includegraphics[width=0.475\textwidth,trim={0 0 0 0},clip]{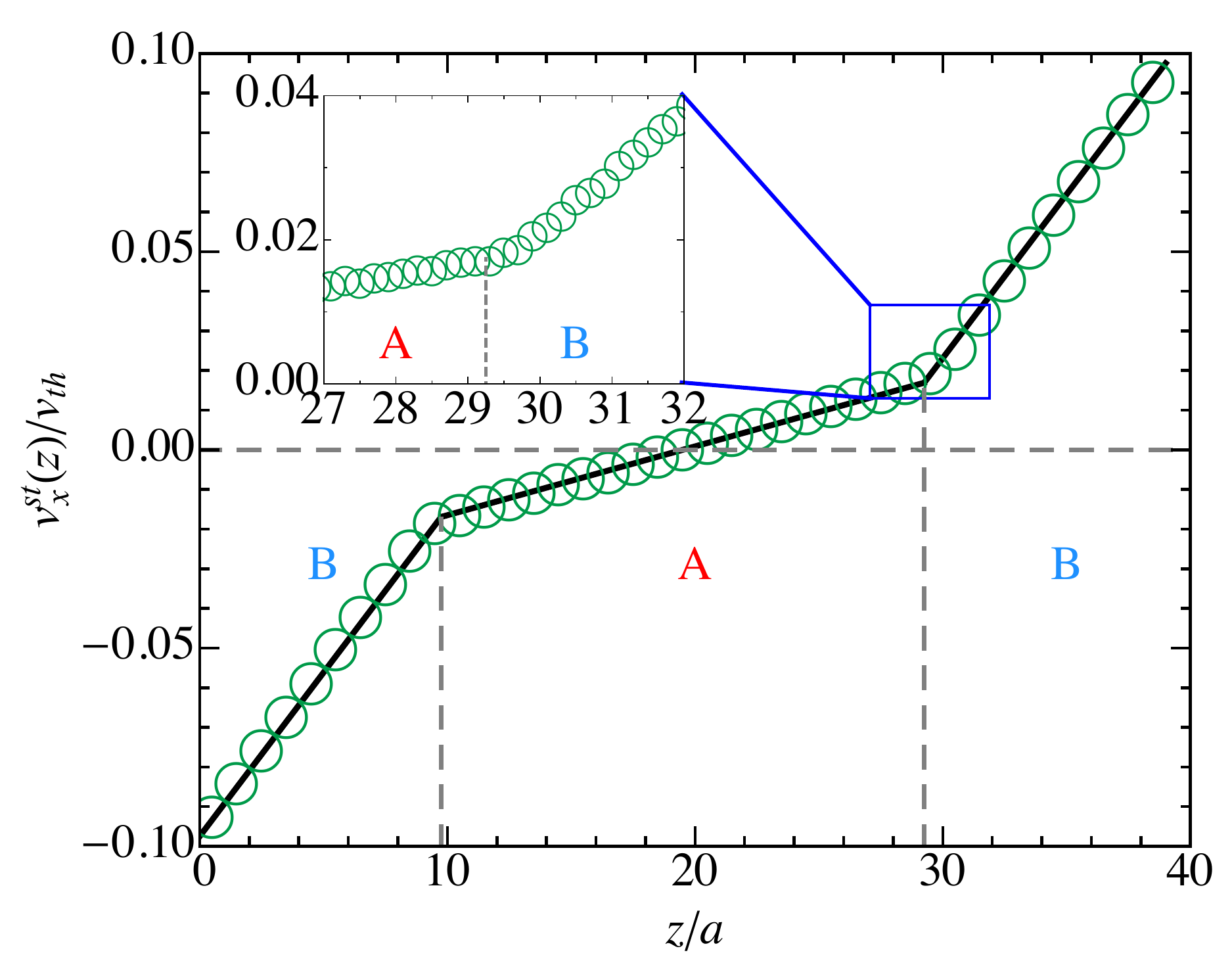}}
			\put(140,15){(a)}
		\end{picture}
		\label{fig:vx}
	}\\[1ex]
	\vspace{0.5cm}
	\subfloat{		
		\begin{picture}(100,180)
			\put(-75,-10){\includegraphics[width=0.48\textwidth,trim={0 0 0 0},clip]{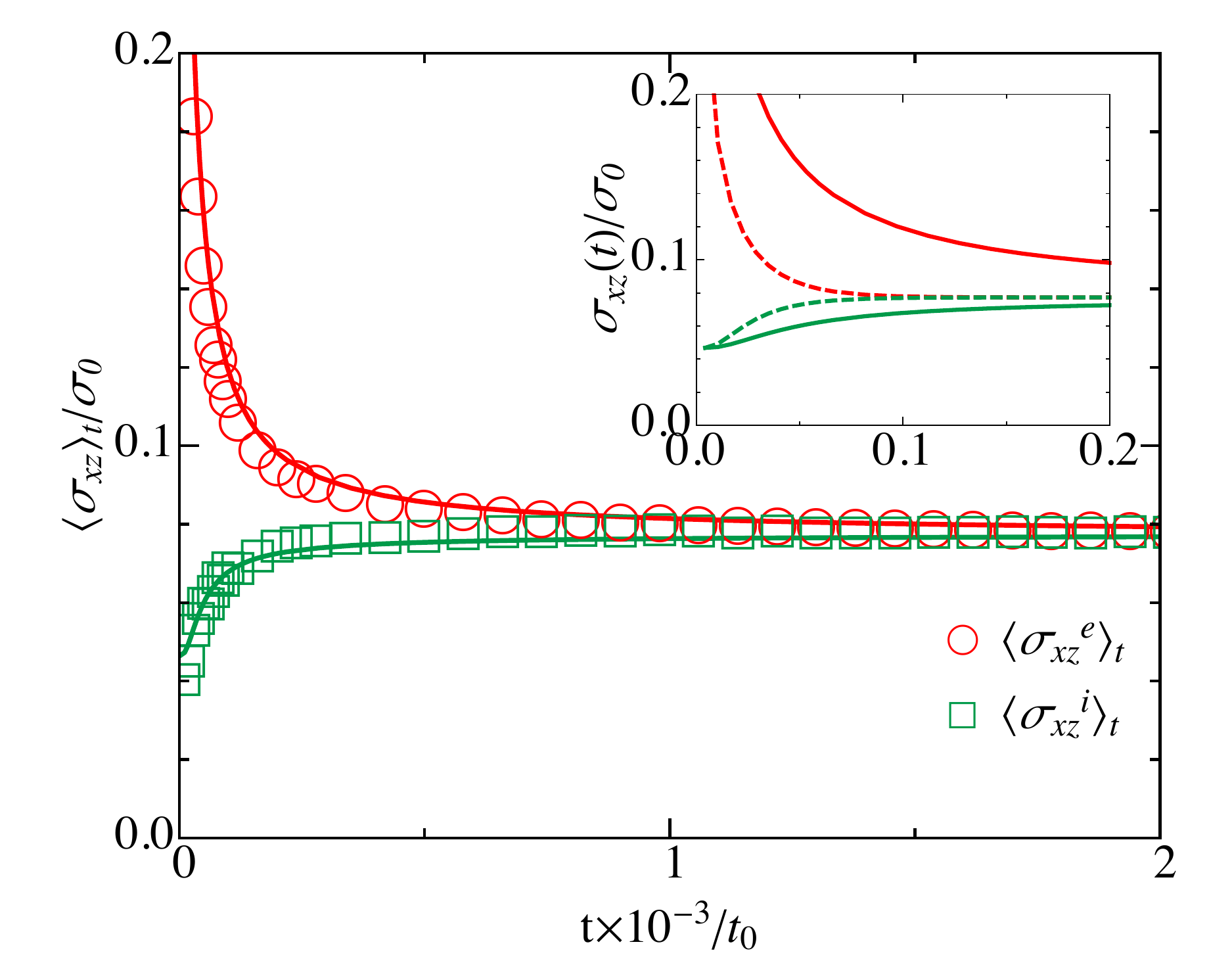}}
			\put(140,20){\textcolor{black}{(b)}}
		\end{picture}
		\label{fig:stress}
	}	\vspace{0.2cm}
	\caption {\label{fig:shear} (a) Stationary shear velocity profile, $v_x^{st}(z)$, of a $B$-$A$-$B$ fluid layer system, obtained from MPC simulations (open circles) and analytically from continuum hydrodynamics (solid line)  according to Eqs.~\eqref{eq:bcs} - \eqref{eq:us}, and Eq. (\ref{eq:StationaryFlowField}) in the Appendix. 
		The magnitude of the wall velocity is $|u|=0.0975 v_{th}$.
		Inset: Magnification of $v_x^{st}(z)$ in the $A$-$B$ interfacial region (blue rectangle). (b)  Moving time average of the external shear stress, $\langle \sigma_{xz}^{e} \rangle_t$, at the upper and lower wall, and of the internal shear stress, $\langle \sigma_{xz}^i \rangle_t$, in units of thermal stress $\sigma_0=k_BT/a^3 =13\;\!\sigma_{xz}^{st}$. Open symbols are MPC data, and solid lines our continuum hydrodynamics predictions (cf.  Appendix). Inset: Continuum hydrodynamics results for the  instantaneous (dashed lines) and moving time average (Eq.~\eqref{eq:Cumulative}) (solid lines) external (red) and internal (green) stresses $\sigma_{xz}^{e,i}(t)$ and $\langle\sigma_{xz}^{e,i}\rangle_t$, respectively, as functions of $t$ (as in the main plot).    }
\end{figure}

\noindent where $u_{s}^+$ and $u_{s}^-=-u_{s}^+$ (by symmetry) are the fluid velocities at the upper and lower fluid interfaces, respectively. The interfacial velocities are obtained using the continuity of shear stress across the planar clean interfaces (no Marangoni stress and Laplace pressure), $\eta_A\;\!\dot{\gamma}_A = \eta_B\;\!\dot{\gamma}_B$, 
%
at $z=L_z/4$ and $3L_z/4$, respectively, which yields
\begin{align} \label{eq:us} 
	u_s^{\pm}=\pm\frac{\eta_B}{\eta_B + \eta_A} u\,.
\end{align}
Here, $\dot{\gamma}_A=d u^{A,st}_x(z)/dz$ and $\dot{\gamma}_B=d u^{B,st}_x(z)/dz$ are the constant shear rates in the mid-layer of fluid $A$ and in the two layers of fluid $B$, respectively. The (dynamic) pressure for the unidirectional shear flow is constant throughout the system including the interfaces. The Appendix presents  our analytical continuum hydrodynamic result for the transient starting flow $v_x(z,t)$ of the $B$-$A$-$B$ system which converges to $v_x^{st}(z)$ in the course of time (cf. Fig. \ref{fig:Profiles}).  

The MPC simulation results for $v_x^{st}(z)$ displayed in Fig.~\ref{fig:shear}(a) reflect the hydrodynamically expected behavior of three linear stationary shear flow regions. Even more, the simulation results agree quantitatively with the hydrodynamic flow profile described in Eqs.~\eqref{eq:bcs} - \eqref{eq:us}, and in Eq. (\ref{eq:StationaryFlowField})  of the Appendix. In spite of the non-zero thickness of the interface in the MPC simulations of the  order of the collision cell size $a$, caused by discretization in terms of collision cells and  random shift of the collision cell lattice, the MPC results suggest that  the interface width is of minor relevance for fluid properties on lengths scales significantly larger than $a$. The inset of Fig.~\ref{fig:shear}(a) magnifies the stationary velocity profile in the $A$-$B$ interfacial region. It suggests a continuous change both of $v_x^{st}(z)$ and its slope across the interface. This indicates also a continuous change of the \textit{local} viscosity in the interfacial region caused by the discretization.

In order to confirm the hydrodynamically expected relations between stress, shear rate, and viscosity  in the MPC simulations of the immiscible fluid, we determine  the instantaneous internal (superscript $i$), $\sigma_{xz}^i$, and  external  (superscript $e$), $\sigma_{xz}^e$, shear stresses. The latter is the stress (modulus) exerted by the fluid particles on the walls, while the former one is the volume averaged stress. Explicitly, the  stresses at a given time instant are computed as~\cite{Winkler2009} 
\begin{align} \nonumber
	\sigma_{xz}^e & = \frac{L}{2Vh_{B}}\left(\sum_{i=1}^{N}\Delta p^{u}_{ix}-\sum_{i=1}^{N}\Delta p^{l}_{ix}\right)\\ & + \frac{L}{2Vh_{B}}\left(\sum_{i\in bc}\Delta p^{u}_{ix}-\sum_{i\in bc}\Delta p^{l}_{ix}\right) ,
	\label{eq:sigma_e} \\ \nonumber 
	\sigma_{xz}^i & = -\frac{1}{V}\sum_{i=1}^{N}m\hat{v}_{ix}\hat{v}_{iz}+\frac{2u}{Vh_{B}}\sum_{i=1}^{N}m\hat{v}_{iz} \Delta t_i \\ & \hspace*{-5mm} - \frac{1}{Vh_{A(B)}} \sum_{i=1}^{N}\Delta p_{ix}r_{iz}+\frac{L}{2Vh_{B}}\left(\sum_{i\in bc}\Delta p^{u}_{ix}-\sum_{i\in bc}\Delta p^{l}_{ix}\right)\,,
	\label{eq:sigma_i}
\end{align}
where the sums extend over all $N$ fluid particles inside the simulation box,  and $r_{iz}$ is the position of the MPC particle $i$ along the $z$-axis.
Here, the change in the momentum, $\Delta\bm{p}_i(t)$, of a particle $i$ in a  collision step is given by
\begin{align}
	\Delta\bm{p}_i(t) =m(\vi(t)-\hat{\v}_i(t))\,,
	\label{eq:pm}
\end{align}
where $\hat{\v}_i$ is the particle velocity after streaming and before collision. The superscripts $u$ and $l$ indicate  that the considered quantity is calculated at the upper and lower wall, respectively. Note that Eqs.~\eqref{eq:sigma_e} and \eqref{eq:sigma_i} account also for momentum exchange due to collisions with phantom particles located  inside wall boundary cells (bc). The negative sign in front of the  transversal momentum exchange $\Delta p^{l}_{ix}$ accounts for the negative velocity, $-u$, of the lower wall. In Eq.~\eqref{eq:sigma_i}, the second term on the right-hand side accounts for  the momentum change of $B$ particles ``reflected'' (bounce-back) at a wall in the streaming step,  and  $\Delta t_i$ is the time during which particle $i$ streams in the fluid before colliding with a wall \cite{Winkler2009}.
The internal stress calculation invokes the momentum exchange of fluid $A$ and $B$ particles, described by the third term on the rhs of Eq.~\eqref{eq:sigma_i}. Here,  time averaging is performed separately for each fluid phase, owing to the different collision times $h_A$ and $h_B$. 

In the simulations, ``macroscopic'' stress tensors are calculated via averaging over various realizations (ensemble average) as well as averaging over time. For the latter, we determine the moving time averages  
\begin{align} \label{eq:Cumulative}
	\langle\sigma^{i,e}_{xz}\rangle_t=\frac{1}{t} \int_0^t\!\!dt'\;\!\sigma^{i,e}_{xz}(t')\,,
\end{align}
of external and internal stresses, which yield the stationary state value, $\sigma_{xz}^{st}$, in the limit $t \to \infty$. As a particular case, we analyze the starting flow situation, where at time $t=0$, the two confining walls suddenly start to move oppositely with constant velocities $\pm u\;\!{\bf e}_x$, respectively. Fluid and walls are at rest for $t<0$.

\indent MPC results for the  moving time average external and internal shear stresses are presented in Fig.~\ref{fig:shear}(b), as function of the elapsed time $t$ after the two walls started to move. The MPC results for the modulus of the external stress, $\langle \sigma^e_{xz}\rangle_t$, at the upper and lower wall are  equal within the accuracy of the simulations. The external stress decays monotonically towards the  plateau value $\sigma_{xz}^{st}= 0.077\;\!\sigma_0$ where the steady-state regime is reached, characterized by the fully developed, piece-wise linear shear profile in Fig.~\ref{fig:shear}(a) and an uniform shear stress. Its decay reflects the diffusive broadening of the region of changing fluid velocity near the walls with increasing time, as depicted in Fig. \ref{fig:Profiles} of the Appendix. The moving time averaged internal stress increases instead from its minimal value at $t=0$, where the bulk fluid is still at rest, towards its steady-state value  $\sigma_{xz}^{st}$. The characteristic transition time for the external shear stress relaxation (and internal stress buildup) towards the uniform steady-state value is estimated as $\tau_v = (L_z/4)^2\left(1/\nu_A+1/\nu_B\right) \approx 140\times t_0$, which is the viscous diffusion time across half of the simulation box, $L_z/2$. The MPC moving time averaged external and internal stresses in the sheared $B$-$A$-$B$ system are in excellent agreement with the corresponding hydrodynamic stresses (solid curves), the latter obtained from the according analytic expressions presented in the Appendix.

From the limiting steady-state stress value and the steady-state flow profile, the viscosity values $\eta_A/\eta_0=42.9$ and $\eta_B/\eta_0 = 9.1$ are deduced, which agree within less than $5\%$  with the viscosity values obtained from analytical theory for respective one-component MPC fluids (cf. Sec.~\ref{sec:sim_para}). 
The shear viscosities of the binary fluid model can be easily controlled by a single parameter, namely the collision time step $h$, but there is a continuous viscosity crossover along the MPC interface of thickness comparable to the collision cell. In general, the continuum hydrodynamic behavior is accurately recovered by the MPC simulations for lengths larger than about $2a$ \cite{huang12}. 

The inset in Fig.~\ref{fig:shear}(b) depicts our continuum mechanics results, summarized in the Appendix, for the instantaneous external and internal shear stresses $\sigma_{xz}^{e,i}(t)$  in comparison with the according moving time averages $\langle\sigma_{xz}^{e,i}\rangle_t$. By their definitions, the  moving stress averages approach the common steady-state value more slowly than the stresses themselves. As shown in the Appendix, $\sigma_{xz}^{e}(t)$ and $\langle\sigma_{xz}^{e}\rangle_t$ exhibit the time-dependence $1/\sqrt{t}$ near $t=0$. In contrast, the internal stress and its time averaged counterpart are finite and minimal at $t=0$, with the common value $0.60\;\! \sigma_{xz}^{st}$ according to Eq. (\ref{eq:InternalStressLimit}).

\section{Hydrodynamic correlations: transverse velocity auto-correlation function}
\label{sec:hydro_corr}
Additional insight into the time-resolved hydrodynamic behavior of the MPC fluid is gained by analyzing  the transverse velocity auto-correlation function (TVCF) \cite{huang12} in the various layers.   
For a stationary and isotropic Newtonian fluid in a volume $V$ with periodic boundary conditions, the linearized Landau-Lifshitz  Navier-Stokes equations yields the single-exponentially decaying TVCF in Fourier space \cite{huang12}
\begin{align}  \label{eq:tcvv}
	\left<\u^T(\bm k,t)^T\cdot\u^T(-\bm k,0)\right>=\frac{2k_BT }{\rho V}e^{-\nu k^2 t}\,.
\end{align}
Here, $\bm u^T(\bm k,t)$ is the Fourier-transformed  velocity part perpendicular to the wave vector $\bm k$~\cite{huang12,theers2015}, i.e., $\bm u^T \cdot \bm k = 0$. The brackets denote an equilibrium ensemble average, with the fluid system at rest on hydrodynamic time and length scales. The factor 2 on the rhs accounts for the two independent transversal modes. Owing to isotropy, the TVCF depends only on the modulus $k=|\bm k|$ of the wave vector. Simulation results for the TVCF of a single-phase MPC fluid are in excellent agreement with the above hydrodynamic prediction \cite{kroll2006,huang12,theers2015}.    

To explore thermally induced transverse velocity correlations in our three-layer model of fluids $A$ and $B$, we perform simulations for a cubic simulation box of size $L=80\;\!a$, with periodic boundary conditions in all three Cartesian coordinate directions. The higher-viscosity layer $A$ of width $L/2$ is symmetrically sandwiched  between two fluid-$B$ layers, as illustrated in Fig.~\ref{fig:shear_sch}, but now for a system without shear. The period boundary condition along the $z$-axis implies an alternating pattern of horizontal $A$ and $B$ layers of equal thickness $L/2$. 

We determine the TVCFs of the pure $A$ and $B$ fluids in the three-layer model by considering $N_o$ MPC particles inside an observation cuboid of $z$-thickness $L_o=L/4 =20\;\!a$ and volume $V_o=L\times L\times L_o$, symmetrically located inside the $A$-fluid and $B$-fluid layers, respectively. To explore additionally the influence of the fluid interface, the TVCF for another observation cuboid with smaller vertical width $L_o=10\;\!a$ is determined, with the cuboid symmetrically enclosing the $A$-$B$ interface. The Fourier transform, $\u(\bm k,t)$, of the fluid velocity fluctuations in an observation cuboid is calculated according to
\begin{align}  \label{eq:vkt}
	\u(\bm k,t)=\frac{1}{N_o}\sum_{i=1}^{N_o}  \bm v_i(t)e^{i\bm k\cdot \bm r_i(t)}\,,
\end{align}
where $\bm v_i(t)$ is the velocity of  fluid particle $i$ at position ${\bm r}_i(t)$ inside the considered cuboid. For the cuboid centered around the $A$-$B$ interface, half of the particles summed over are, on average, of $A$-type and half of $B$-type.  For the present purpose, we consider only wave vectors ${\bm k}={\bm k}_\parallel$ parallel to the $xy$-plane, with wavelength $\lambda=2\pi/k$ smaller than the  cuboid width $L_o$. This reduces boundary artifacts due to fluid particles leaving or entering the observation cuboid.

\begin{figure}[H]
	\centering
	\subfloat{		
		\begin{picture}(100,180)
			\put(-78,-15){\includegraphics[width=0.48\textwidth,trim={0 0 0 0},clip]{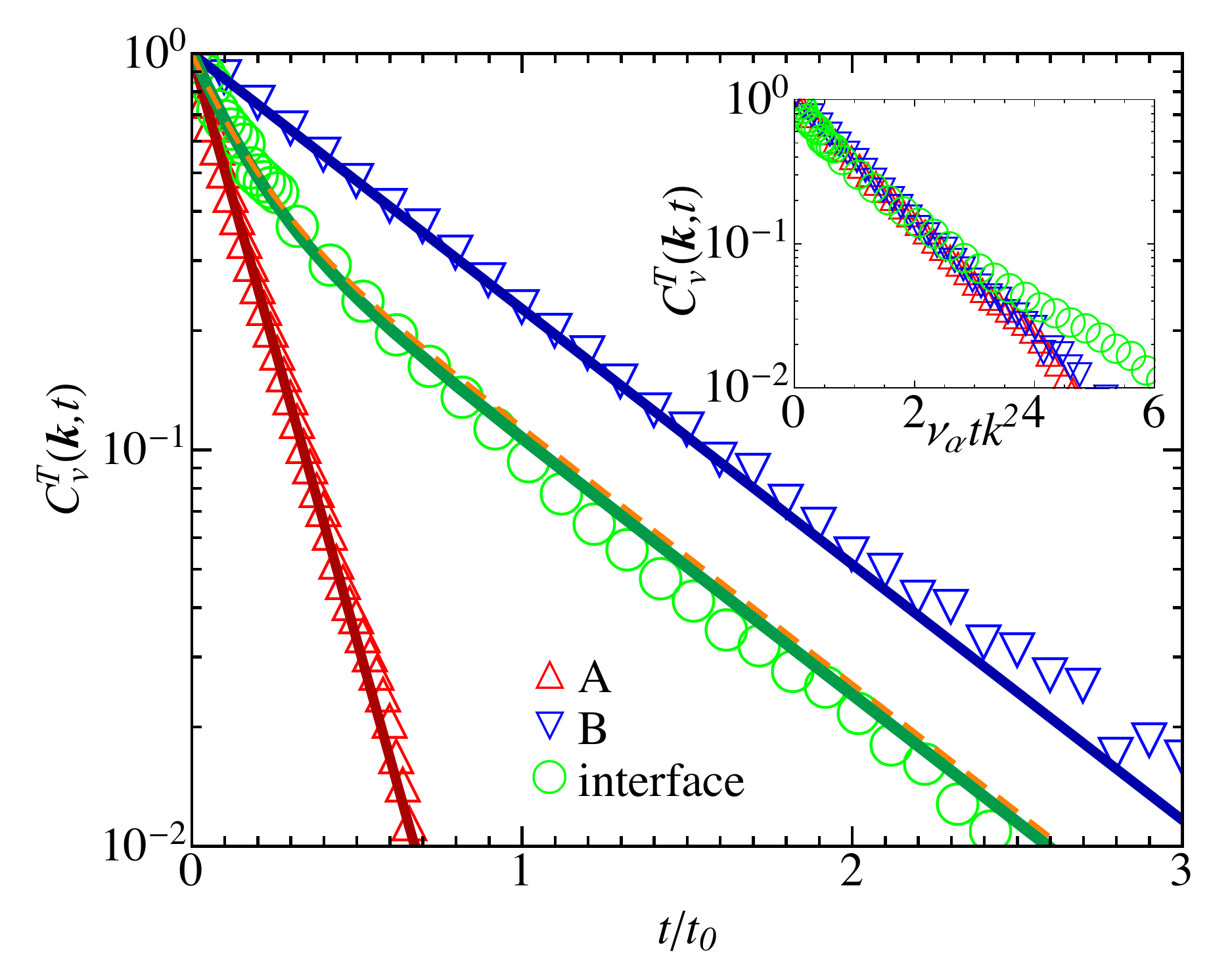}}
			\put(-35,15){(a)}
		\end{picture}
		\label{fig:tcvv}
	}\\[1ex]
	\vspace{0.5cm}
	\subfloat{		
		\begin{picture}(100,180)
			\put(-78,-10){\includegraphics[width=0.48\textwidth,trim={0 0 0 0},clip]{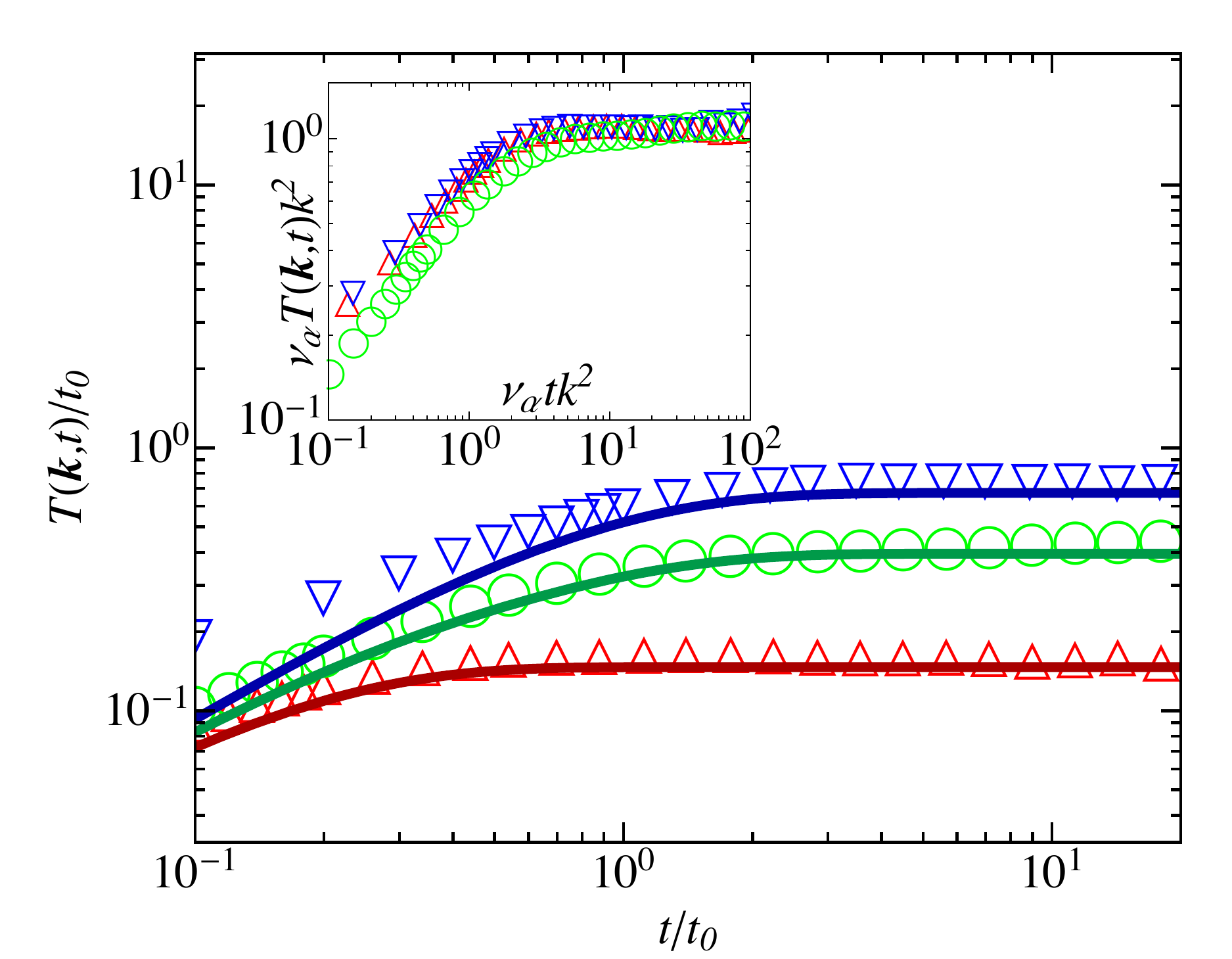}}
			\put(-35,18){\textcolor{black}{(b)}}
		\end{picture}
		\label{fig:TK}
	}
	\vspace{0.2cm}
	\caption{\label{fig:tcvvTK}  (a)~Normalized TVCFs, $C_v^T({\bm k},t)$,  for a wave vector ${\bm k}$ parallel  to $xy$-plane with wavelength $\lambda= 5a$ obtained from MPC simulations for the fluid $A$ (red, up triangles), $B$ (blue, down triangles), and the  $A$-$B$ interface cuboid (green, circles). The corresponding solid lines represent the bulk-fluid prediction $\exp(-\nu_{A,B} k^2t)$ and the double-exponential expression in Eq.~\eqref{eq:doppExp} with weights $\xi_A=0.541=1-\xi_B$ (green solid line) and $\xi_A=\xi_B=1/2$ (orange dashed line), respectively. Inset: TVCF dependence on $\nu_\alpha k^2 t$ for $\alpha \in \{A,B,\text{int}\}$ and $\nu_\text{int}=\nu_A\nu_B/[\xi_A\nu_A+\xi_B\nu_B]$.
		(b)~Time-integrated normalized TVCFs, $T(\bm k,t)$ \eqref{eq:Tk}. Solid lines represent Eq.~\eqref{eq:Tk} and \eqref{eq:doppTk}, respectively. Inset: data collapse for  $\nu_\alpha k^2 t$.}
\end{figure}

\indent Figure~\ref{fig:tcvvTK}(a) displays the normalized TVCFs
\begin{align} 	\label{eq:ntcvv}
	C_v^T({\bm k},t)=\frac{\left<\u^T({\bm k},t)\cdot\u^T(-{\bm k},0)\right>}{ \left<\u^T({\bm k},0)\cdot\u^T(-{\bm k},0)\right>} 
\end{align}
of the fluid inside the $A$ and $B$ cuboids, respectively, as well as the TVCF of the mixed-fluid cuboid enclosing the $A$-$B$ interface. The horizontally oriented wavectors employed here are ${\bm k} = (32\pi/L)(1,0,0)$ and ${\bm k} = (32 \pi/L)(0,1,0)$, of wavelength $\lambda= 5 a$ smaller than the cuboid width. Notice that $\u^T({\bm k},t)=({\bm 1}-{\bm k} \otimes	 {\bm k}/k^2) \u({\bm k},t)$.  Owing to the non-isotropic three-layer structure, $C_v^T({\bm k},t)$ is in principle an anisotropic function of the wave vector, depending also on the vertical location and width of the considered cuboid.

Within the correlation time window $t \leq 3 t_0$ depicted in Figure~\ref{fig:tcvvTK}(a), the MPC-calculated  normalized TVCFs of the pure $A$ and $B$-fluid cuboids (open symbols) decay exponentially according to $\exp(-k^2 \nu t)$, with kinematic viscosity values $\nu_A$ and $\nu_B$ as numerically obtained in Sec.~\ref{sec:shear}. The reason why the isotropic bulk fluid TVCF form is recovered in the anisotropic three-layer system (within numerical accuracy) is that the viscous diffusion time, $ \tau^{A,B}_\nu =(L/8)^2/\nu_{A,B}$, over a distance from the cuboid center to the interface is large compared to the resolved correlation time window; the viscous diffusion times are $\tau^{A}_\nu = 24 t_0$ and $\tau^{B}_\nu =115 t_0$, respectively. Hence, in the considered time window and the considered wave vector of wavelength $\lambda = 5a$, the velocity correlations in the single-fluid cuboids are yet unperturbed by the interfaces.  

On the same basis one could expect that the MPC data for the TVCF of the cuboid symmetrically enclosing the $A$-$B$ interface are for $t \leq 3 t_0$ decently well reproduced by the superposition of two bulk-fluid exponential TVCFs according to 
\begin{align}  \label{eq:doppExp}
	C_v^{T,int}(\bm k,t)=\xi_A e^{-\nu_A k^2 t}+\xi_Be^{-\nu_B k^2 t}\,,
\end{align}
for equal weight factors $\xi_A=\xi_B =1/2$, and $\nu_A$ and $\nu_B$ determined from the MPC simulation data for the single-fluid layers. The equal-weight superposition according to Eq.~\eqref{eq:doppExp}  somewhat underestimates the decay of the correlation function for  $t >  t_0$, reflecting the growing influence of the interfacial region with increasing time. A fit of the simulation data by Eq.~\eqref{eq:doppExp}, for unchanged values of $\nu_A$ and $\nu_B$, yields the  weight factors $\xi_A=0.541$ and $\xi_B=1-\xi_A=0.469$ (dark-green solid line). The asymmetry could be a consequence of the shorter viscous diffusion time across the half-width $L_o/2$ for the fluid-$A$ part of the two-fluid observation cuboid.

The time integral of the normalized TVCF \eqref{eq:ntcvv}, characterizing a one-component fluid in the hydrodynamic regime, is
\begin{align} \label{eq:Tk} 
	T(\bm k, t)=\int_0^t\!\!dt'\;\!C_v^T(\bm k,t')=\frac{1}{\nu_\alpha k^2}\left(1-e^{-\nu_\alpha k^2 t}\right) ,
\end{align}
where $T(\bm k,t)$, in the limit $t\to \infty$, is related to the  Oseen tensor in reciprocal space \cite{dhont1996introduction,doi1988theory,huang12}.   
The accordingly time-integrated TVCF in Eq.~\eqref{eq:doppExp} for the cuboid enclosing the interfacial region is
\begin{align} \label{eq:doppTk}
	T^{int}(\bm k, t)=&\frac{\xi_A }{\nu_A k^2}\left(1-e^{-\nu_A k^2 t}\right)+\frac{\xi_B }{\nu_B k^2}\left(1-e^{-\nu_B k^2 t}\right)\,,
\end{align}
with $T^{int}(\bm k,\infty)=1/ k^2\nu_{int}$ and $\nu_\text{int}=\nu_A\nu_B/[\xi_A\nu_A+\xi_B\nu_B]$. 

The time dependence of $T(\bm k,t)$ and $T^{int}(\bm k, t)$ for the three observation cuboids, obtained from the data of Fig.~\ref{fig:tcvvTK}(a), are shown in Fig~\ref{fig:tcvvTK}(b). The time-integrated MPC simulation data (open symbols) agree overall well with the analytic  expressions in Eqs.~\eqref{eq:Tk} and \eqref{eq:doppTk} based on the single-fluid theoretical expressions. As shown in the inset,  the time-integrated TVCFs are universal functions of $\nu_\alpha k^2 t$, as expected by the identical universal behavior of the TVCFs. The factor  $\nu_{\text{int}}=\nu_A\nu_B/[\xi_A\nu_B+\xi_B\nu_A]$ can be considered as a common effective kinematic viscosity of the $A$ and $B$ fluid contributions in the cuboid enclosing the interface. The inset further illustrates that the crossover to the long-time plateau values $1/(k^2\nu_\alpha)$ is characterized  by the viscous diffusion times $\tau^\alpha_k=(\nu_\alpha k^2)^{-1}=0.63 a^2/\nu_\alpha$. Since $\tau^A_k<\tau^{\text{int}}_k<\tau^B_k$, the Stokesian regime of inertia-free, quasi-instantaneous hydrodynamics is reached for the considered wavenumber at times distinctly smaller than the viscous diffusion time across the colloid diameter $(2R)^2/\nu_B=29t_0 \gg \tau_k^B$.

\section{Mobility of a colloidal sphere near a fluid-fluid interface} \label{sec:mobility}

\begin{figure}[ht]
	\vspace{1.0cm}
	\hspace{-0.9cm}
\subfloat{		
	\begin{picture}(100,100)
		\put(-20,5){\includegraphics[width=0.14\textwidth,trim={0 0 0 0},clip]{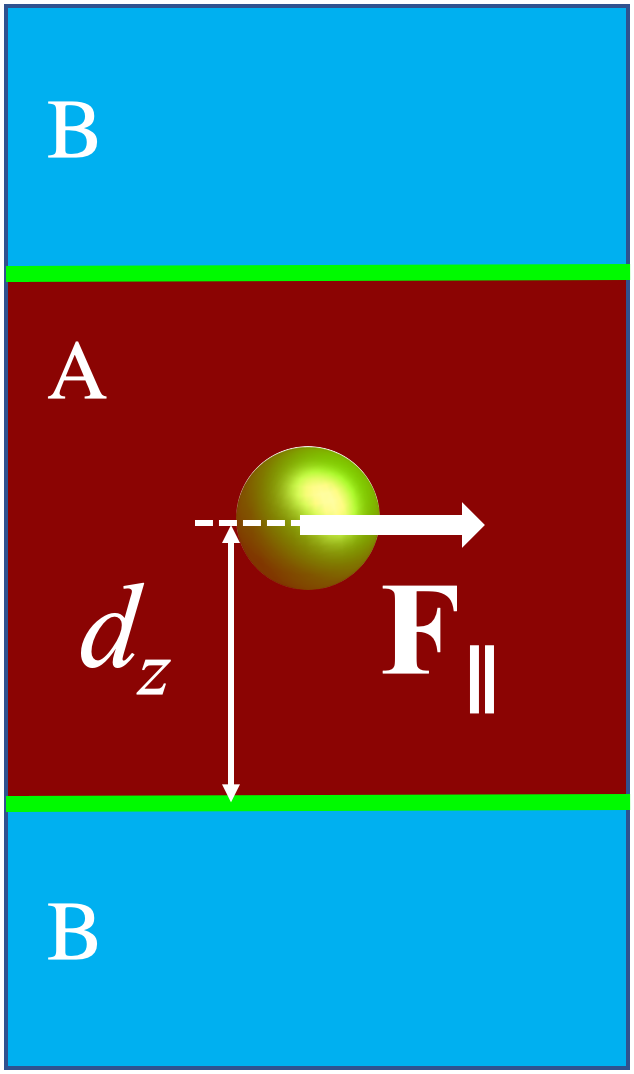}}
		\put(-7.0,114){(a)}
	\end{picture}
	\label{fig:lmobility_scheme}
}
\hspace{-1.1cm}
\subfloat{		
	\begin{picture}(100,100)
		\put(-20,-5){\includegraphics[width=0.31\textwidth,trim={0 0 0 0},clip]{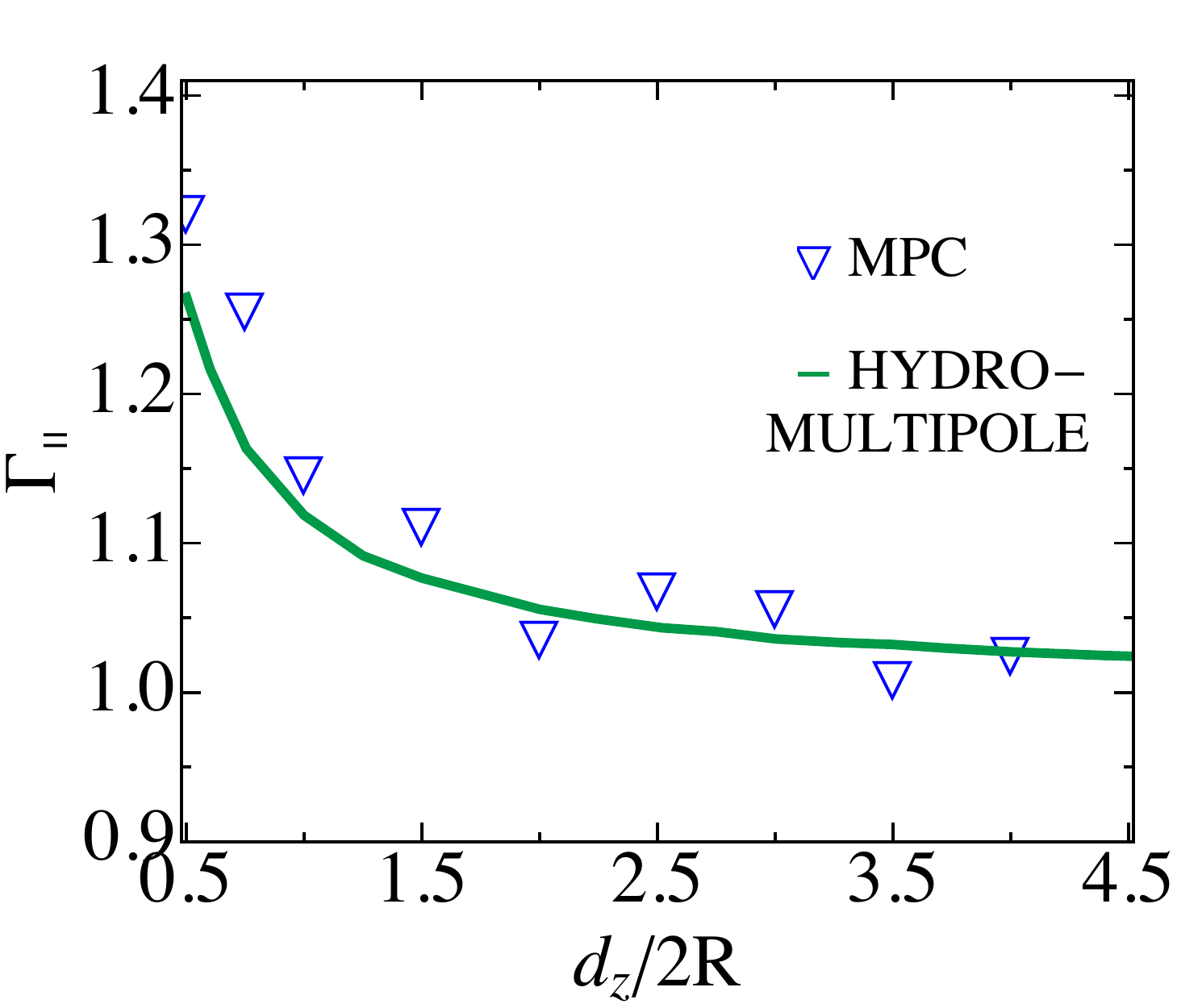}}
		\put(10.0,110){(c)}
	\end{picture}
	\label{fig:lm}
}
\vspace{1.0cm}
\subfloat{		
	\begin{picture}(100,100)
		\put(-33,5){\includegraphics[width=0.14\textwidth,trim={0 0 0 0},clip]{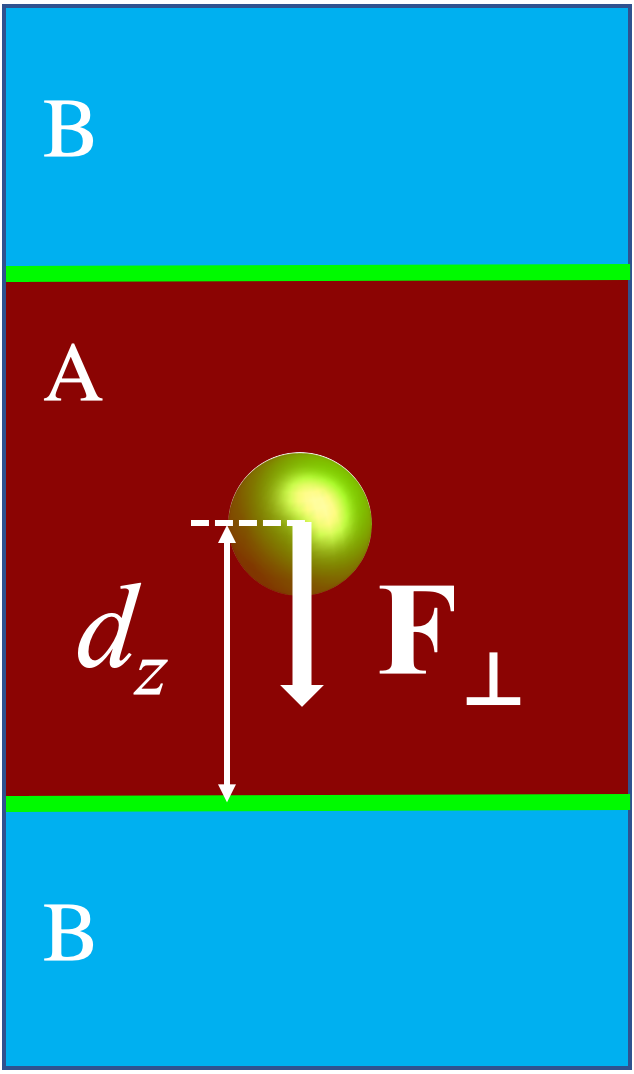}}
		\put(-20.0,114){(b)}
	\end{picture}
	\label{fig:tmobility_scheme}
}
\hspace{-1.2cm}
\subfloat{		
	\begin{picture}(100,100)
		\put(-30,-5){\includegraphics[width=0.31\textwidth,trim={0 0 0 0},clip]{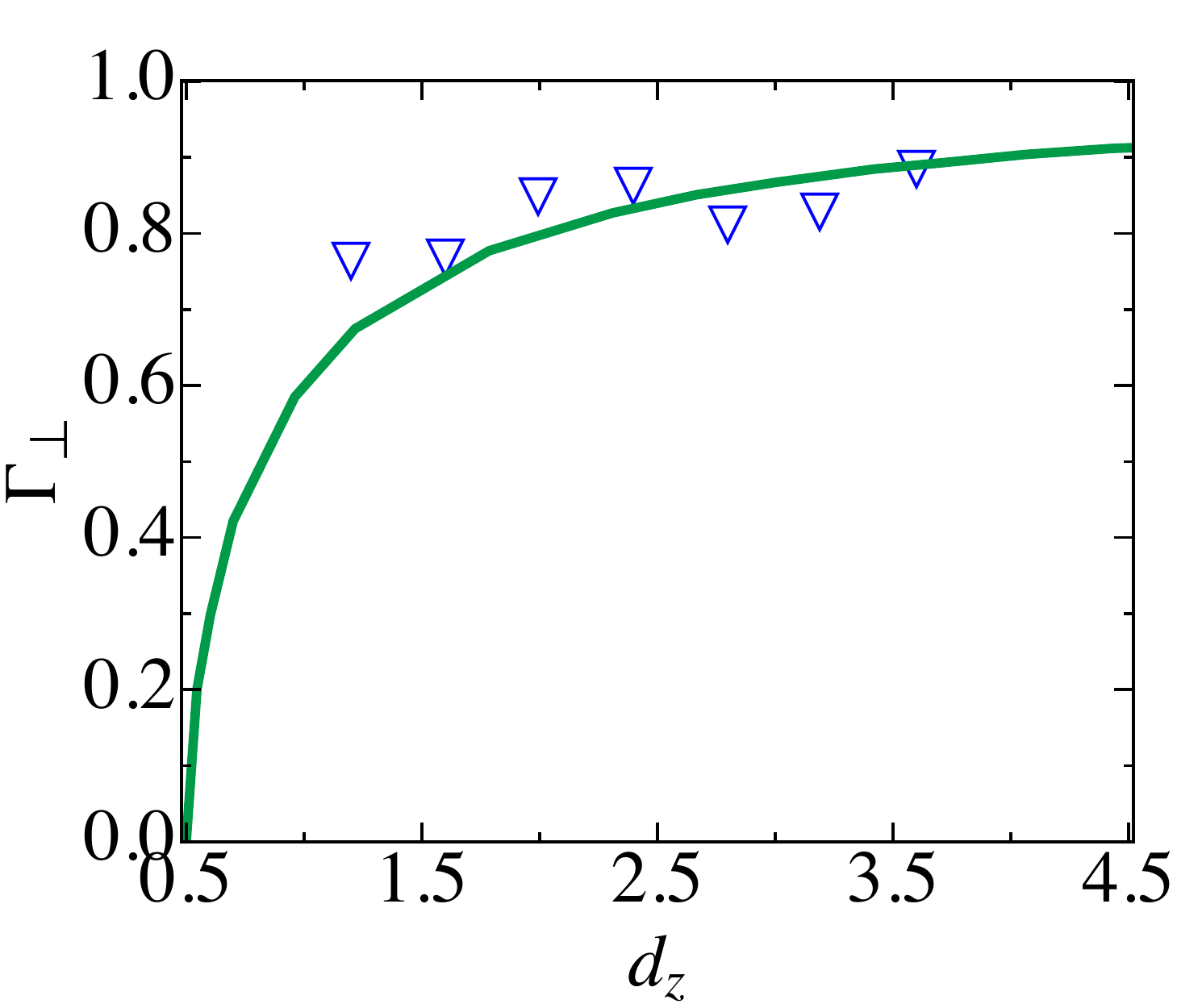}}
		\put(0.0,110){(d)}
	\end{picture}
	\label{fig:tm}
}
	\caption {\label{fig:mpcpmc} (a) and (b) Schematics of the three-layer system for determining the lateral and transverse translational mobility coefficients, $\Gamma_\parallel$ and $\Gamma_\perp$, of a no-slip sphere of radius $R=2.5 a$ embedded in fluid $A$, as functions of the reduced distance $d_z=z/(2R)$ from the sphere center to the $A$-$B$ interface. The employed viscosity ratio is $\eta_B/\eta_A= 0.21$. (c) and (d)  Lateral and transverse mobility coefficients from MPC simulations (symbols)  and hydrodynamic force-multipole-expansion calculations (green lines)~\cite{blawzdziewicz2010} in units of the bulk mobility value $1/(6 \pi \eta_A R)$.}
\end{figure}
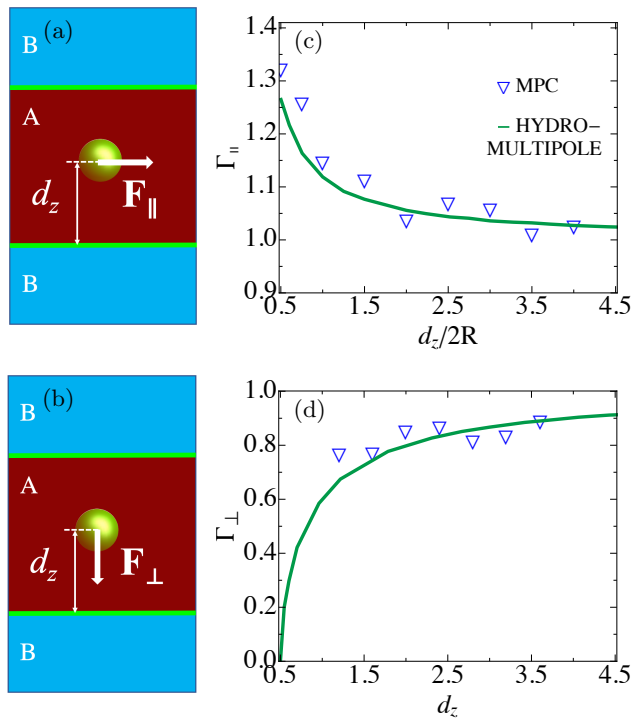

The (strong) viscosity difference between two immiscible fluid phases affects the dynamics of a colloidal particle moving near the fluid interface. To explore the hydrodynamic coupling, and to scrutinize the according MPC coupling predictions in our three-layer model, we calculate the mobility coefficients of a sphere embedded in fluid $A$ which moves steadily under low-Reynolds-number conditions parallel or perpendicular to the planar $A$-$B$ interfaces. The coefficients are determined as functions of the reduced distance $d_z=z/(2R)$ of the sphere center from the  $A$-$B$ interface (see Fig.~\ref{fig:mpcpmc}). To reduce finite-size effects due to the periodic boundary conditions in $z$-direction, different from Secs.~\ref{sec:hydro_corr}, we consider a non-cubic  simulation box of lengths $2L_x=2L_y=L_z=80a$. 

As indicated in Fig.~\ref{fig:mpcpmc}, the sphere of radius $R=2.5a$ is subjected to a weak constant force $F_\parallel$ ($ F_\perp$) applied to its center, and oriented parallel (perpendicular) to the fluid-fluid interface.   Due to the no-slip boundary conditions employed on the sphere surface, the moving sphere drags nearby fluid along, which is compensated  by fluid backflow such that the total momentum of the system in any spatial direction is zero (quiescent fluid system assumed) \cite{padding2004,singh2018}. Under low-Reynolds-conditions, where in the continuum mechanics picture the fluid flow is described by the quasi-stationary linear Stokes equation, the reduced translational mobilities follow from the relations 
\begin{align} \label{eq:mobilities}
	\Gamma_\parallel(d_z)&=6\pi\eta_AR\;\!\frac{\langle v_\parallel (d_z) \rangle}{F_\parallel} \, ,\\
	\Gamma_\perp(d_z)&=6\pi\eta_AR\;\!\frac{\langle v_\perp (d_z) \rangle}{F_\perp} 
\end{align}
by measuring the steady-state mean velocity $\langle v_{\parallel,\perp}\rangle$ of the sphere for a given constant force $F_{\parallel,\perp}$. In our MPC simulations, the thermal force value $F_{\parallel,\perp}=4k_BT/a$ is used. After applying the force to the sphere, the steady-state with constant mean drift velocity is reached for times $t \gg R^2/\nu_A$. Note that $\Gamma_{\parallel,\perp}=1$ in the bulk region of the fluid far from any interface or boundary \cite{dhont1996introduction}. 

Figures~\ref{fig:mpcpmc}(c) and (d) display the MPC results (open blue symbols) for the normalized lateral and transverse mobilities, $\Gamma_\parallel(d_z)$ and $\Gamma_\perp(d_z)$, as functions of the reduced distance $d_z=z/(2R)$. For comparison, according reduced mobilities are shown (green solid lines) as obtained numerically using an elaborate Stokesian dynamics-based hydrodynamic force-multipole expansion method, encoded in the software package HYDROMULTIPOLE \cite{blawzdziewicz2010}. The depicted mobility curves by the force multipoles method, valid under creeping-flow conditions, are taken from \cite{blawzdziewicz2010} and constitute accurate continuum hydrodynamics results for a no-slip sphere in a half-infinite Newtonian fluid $A$ which moves steadily parallel or perpendicular to an ideally flat and clean interface of zero interfacial viscosity and Marangoni stress. The interface separates the fluid-$A$ half-space from the fluid-$B$ half-space. Note that $\Gamma_{\parallel,\perp}$ depend on the ratio of the shear viscosities of the two fluids.      

Both the MPC simulation and continuum hydrodynamic results predict the lateral sphere mobility to increase with decreasing distance $d_z$ from the $A$-$B$ interface. They are in good overall agreement, except for small distances where the simulation data are somewhat larger (see Fig.~\ref{fig:mpcpmc} (c)). Regarding the transverse mobility depicted in Fig.~\ref{fig:mpcpmc} (d), the continuum hydrodynamics curve for $\Gamma_\perp$ decreases strongly with decreasing distance, and assumes the  value  $\Gamma_\perp=0$ at the sphere-interface contact distance $d_z=0.5$ due to lubrication. In contrast, while the MPC simulation data in Fig.~\ref{fig:mpcpmc} (d) are in accord with a mild decline of the mobility for deceasing distance $d_z \gtrsim 1.5$, they do not reproduce the strong drop in $\Gamma_\perp$ at small distances $d_z \lesssim 1$ (i.e., $z \lesssim 5a$). On first sight, this discrepancy is surprising, since friction and lubrication effects for a hard-sphere colloid embedded in a single MPC fluid close to a {\em solid} no-slip wall are well reproduced \cite{padd:10}. However, it can be attributed qualitatively to the mixing of the two fluids in the interfacial region over a thickness larger than a collision cell size $a$, and to a local perturbation of the hydrodynamic flow field by the no-slip sphere moving normally to the nearby interface. In the HYDROMULTIPOLE calculations, the two fluid half-spaces are taken as ideally incompressible, and the interface as ideally thin and flat, without any sphere-induced perturbation. Notice further that the sphere size is comparable with the MPC interfacial thickness. At any rate, the MPC implementation of immiscible fluids captures the dynamics of the immersed colloidal sphere overall quite well.  

\section{Summary and Conclusions}
\label{sec:suc}

In this article, we have presented a MPC-based mesoscale hydrodynamic simulation scheme for modeling immiscible (layered) binary fluids with viscosity contrast  separated by a flat interface.

Shear flow, external and internal shear stress, fluctuating hydrodynamic velocity correlations, and the  hydrodynamic mobilities of an embedded spherical particle moving close to a flat fluid interface have been analyzed for a three-layer MPC fluid, and validated against continuum hydrodynamics predictions. We obtained a piece-wise linear stationary fluid velocity profile in excellent agreement with the continuum hydrodynamics prediction. By computing the shear stress in relation to the shear rate, we confirmed that the analytically obtained viscosity values for single-phase MPC fluids are reproduced by the binary fluid model, in regions distant from the fluid-fluid interface.  Considering the build-up of the shear profile from a resting fluid, we obtain  excellent agreement  between  MPC simulation results and hydrodynamic predictions for the moving time averages of the external and internal shear stresses, in the time range assessed in the simulations. For this comparison, we have derived an analytical solution of the  linearized Navier-Stokes equation  for the shear stress functions of the $B$-$A$-$B$ system under starting flow conditions, using  the Laplace transformation technique. The analytic expressions have allowed us  to assess quantitatively the differences between instantaneous and moving time averaged  stress functions.    

To examine the predictions by our two-fluids MPC model regarding time-dependent correlations of thermally induced velocity fluctuations, we calculated the transverse velocity auto-correlation function (TVCF) in different observation cuboids. We showed that the TVCFs for the single-fluid cuboids follow closely the expected exponential decay, characterized by the kinematic viscosity of the respective fluid and the considered wave number. In contrast, the calculated TVCF of the cuboid enclosing the $A$-$B$ interface is overall well fitted by a linear combination of the exponential TVCFs for bulk fluids $A$ and $B$, using the viscosity values determined in our shear-flow  studies. The approximate validity of linear superposition suggests that the TVCF of the cuboid is only mildly affected by the interfacial region. A stronger interfacial influence can be expected for a narrower cuboid of width smaller than the  employed value $L_o=10a$.

Finally, we have probed the hydrodynamic coupling of a steadily  moving  no-slip sphere to a nearby  flat  two-fluids  interface  by determining its hydrodynamic mobilities. The distance dependence of the  lateral mobility coefficient for the three-layers MPC model agrees well  with the according mobility result by a hydrodynamic force-multipole expansion method for a sphere moving close to an ideally flat, clean interface separating two incompressible fluids. While decent agreement is observed also regarding the transverse mobility for sphere-interface distances larger than three times the sphere radius, the sharp mobility decline at small distances predicted by the continuum hydrodynamics approach for a non-deformable planar interface of zero thickness, is not obtained  by the MPC simulations. We attribute this to the mixing of the two fluids in the MPC interfacial region of thickness larger than the collision cell size $a$, and possibly a local perturbation of the interface caused by the transverse motion of the sphere. Moreover, and different from what is assumed in the force multipoles calculation, the two fluids in the MPC model are compressible. The non-zero compressibility of the fluids may play a role in particular for transverse (i.e., squeezing) sphere motions. To reduce the influence of the finite interface width on the sphere mobilities, a significantly larger sphere can be considered. Moreover, an alternative method to determine the mobilities may be better suited \cite{padd:10} in order to reduce fluid perturbations by the translating sphere. Here, further studies will be performed in the future.

A numerical advantage of the two-fluids MPC model is that the desired viscosities of the fluid phases can be easily prescribed using the analytic viscosity expression for a single-phase MPC fluid \cite{huang12}. Compared to other mesoscale simulation models of immiscible binary fluids, the present model is straightforwardly implemented, since it does not involve the computation of thermodynamic properties and kinetic processes related to phase separation. Hence, the computational cost is comparable to simulating two single-phase MPC fluids with different collision times.

The two-fluids MPC simulation method can be applied to a wide range of biological soft matter systems. For example, the approach can be suitably extended to study interfacial rheological properties including interfacial viscosity ~\cite{Shkulipa2005,denOtter2007} and interfacial tension.  
Furthermore, as noted already in the introduction, the model can be applied to investigate the lateral self- and collective diffusion of different in-membrane or membrane-attached proteins. The effects of the viscosity contrast between a membrane and the adjacent cytosol, and hydrodynamic interactions between proteins and membrane, and among the proteins, on protein diffusion can be simulated over several timescales using a simplifying coarse-graining of the system. In this context, we will perform further simulations, specifically of colloids with a diameter comparable to the width of the fluid $A$ domain. In a more refined analysis, lipid molecules and other macromolecules forming the membrane constitute a crowded environment which slows down the diffusion of embedded proteins~\cite{enge:2005}. Molecular crowding effects cause so-called sub-diffusion, identified recently to play a  vital role in many biological phenomena~\cite{mint:2000,skim:2009,wang:2012,mats:2014}, including neuronal signaling~\cite{hell:2012,sere:2012}. For future assessment, molecular crowding mechanism can be implemented into our three-layer model, with the middle layer playing the role of the membrane, by adding a planar layer of interacting host particles to the middle layer, or alternatively and more realistically, by accounting for visco-elastic effects in the middle layer through semi-atomistic memory function calculations. Work by us in both directions is in progress.

Furthermore, the present MPC model can be extend to interfaces with imposed sinusoidal fluctuations mimicking membrane fluctuations. Moreover, the quantitative control of viscosity values opens the possibility to study systems with designed viscosity gradients. This provides a means to study the dynamics of biological macromolecules or microorganisms responding to viscosity gradients, such as in viscotaxis~\cite{liebchen2018}.

\section*{Acknowledgments}
ZT thanks Marisol Ripoll (Forschungszentrum J\"ulich) for valuable discussions. The authors gratefully acknowledge computing time granted through JARA-HPC on the supercomputer JURECA at Forschungszentrum J\"ulich~\cite{jureca}.  This work is embedded in the joint IBI-INM project on mesoscale modeling of intracellular signaling events underlying neuron functions.
\section*{Appendix}
\label{sec:app}

We present here the analytic hydrodynamic expressions for the transient (starting flow) shear stress and velocity profiles for the layered two-fluid system depicted in Fig. \ref{fig:Sc}, valid under laminar flow conditions where the linearized Navier-Stokes equation of continuum hydrodynamics applies to.

Consider the horizontal planar walls at $z=\pm L_z/2$ to be quasi-instantaneously set into motion at time $t=0$, with constant velocities $\pm u\;\!{\bm e}_x$,respectively. The $B$-$A$-$B$ layered composite fluid, originally at rest for $t<0$, is then gradually set into laminar motion, through the diffusive transport of momentum (viscous stress) away from the moving walls into the fluid interior. The accordingly unidirectional starting flow velocity field, ${\bm v}(z,t)=v_x(z,t){\bm e}_x$, is incompressible. Since the fluid pressure remains spatially uniform,  the linearized  Navier-Stokes equation for the starting flow reduces to the one-dimensional transversal momentum diffusion equations 
\begin{align} \label{eq:Stokes}\nonumber
	\frac{\partial v_x^B(z,t)}{\partial t }=\nu_B\frac{\partial^2 v_x^B}{\partial z^2 }\,,\; 0< z < L_z/4 , \\
	\frac{\partial v_x^A(z,t)}{\partial t }=\nu_A\frac{\partial^2 v_x^A}{\partial z^2 }\,,\; L_z < z < L_z/2 
\end{align}
for fluids $A$ and $B$, respectively. For convenience, $z=0$ denotes throughout the Appendix the midplane between the two walls in Fig. \ref{fig:Sc}, so that $v_x(-z,t) =-v_x(z,t)$ by symmetry. The initial and boundary conditions are here
\begin{align} \label{eq:BCs}\nonumber
	v_x^{A,B}(z,t=0)&=0 , \\ \nonumber
	v_x^A(z=0,t)&=0 , \\ \nonumber
	v_x^B(z=L_z/2,t)&=u , \\ \nonumber
	v_x^A(z=L_z/4,t)&=v_x^B(z=L_z/4,t) , \\ 
	\eta_A \frac{\partial v_x^A}{\partial z}(z=L_z/4,t)&=\eta_B \frac{\partial v_x^B}{\partial z}(z=L_z/4,t) ,
\end{align}
expressing that fluid A sticks to the walls, and that velocity and shear stress are changing continuously across the A-B fluid interfaces at $z=\pm L_z/4$. 

We have solved this linear boundary value problem using the time Laplace transform method and the method of residues, to obtain the composite velocity fields of $A$ and $B$ 
in the time domain. Skipping the lengthy derivation, we directly present the infinite series solution 
\begin{align} \label{eq:FlowField}
	v_x(\overline{z},\tau)=v_x^A(\overline{z},\tau) \chi_A(\overline{z}) + v_x^B(\overline{z},\tau) \chi_B(\overline{z})\,,
\end{align}
with
\begin{widetext}
	\begin{align} \label{eq:FlowFieldParts}\nonumber
		\frac{v_x^A(\overline{z},\tau)}{u} = & \frac{v^{A,st}_x(\overline{z})}{u}+\sum_{k=1}^\infty \frac{4\;\!\mu }{w_k}\frac{\sin(\frac{\overline{z}\;\!w_k}{\mu})}{N(w_k) }\;\! \exp (-w_k^2\;\!\tau )  ,\\ 
		\frac{v_x^B(\overline{z},\tau)}{u} = & \frac{v^{B,st}_x(\overline{z})}{u}+ \sum_{k=1}^\infty \frac{4\;\!\mu }{w_k} 
		\frac{ \cos\left[w_k(\overline{z}-\frac{1}{2})\right] \sin(\frac{w_k}{2\;\!\mu})+\mu \sin\left[w_k(\overline{z}-1/2)\right]\cos(\frac{w_k}{2\;\!\mu})}{N(w_k)} \exp ( -w_k^2\;\!\tau )
	\end{align}
\end{widetext}
and denominator function
\begin{align}\label{eq:Denominator}
	N(w)&=(1+\mu^2)\cos(\frac{w}{2})\cos(\frac{w}{2\mu})-2\;\!\mu \sin(\frac{w}{2})\sin(\frac{w}{2\mu})\,.
\end{align}
Here, $\mu^2 = \nu_A/\nu_B =\eta_A/\eta_B$, since we have assumed equal mass densities of both fluids such as in the MPC simulations. Moreover, $\overline{z}= 2z/L_z$ is the reduced vertical distance and $\tau = 4\;\!t \nu_B/L_z^2$ the reduced time. The characteristic functions for the considered slabs $[0,L_z/4]$ and $[L_z/4,L_z/2]$ of $A$ of $B$ fluids, respectively, are $\chi_A(\overline{z})= \Theta(1/2-\overline{z})\Theta(\overline{z})$ and $\chi_B(\overline{z})= \Theta(1-\overline{z})\Theta(\overline{z}-1/2)$, respectively, with $\Theta$ denoting the unit step function. 

The piecewise linear, long-time stationary velocity field has the fluid $A$ and $B$ contributions
\begin{align} \label{eq:StationaryFlowField} \nonumber
	\frac{v^{A,st}_x(\overline{z})}{u}&=\frac{2\;\!\overline{z}}{1+\mu^2} , \\
	\frac{v^{B,st}_x(\overline{z})}{u}&=\frac{1}{1+\mu^2}\left[1+2\mu^2\left(\overline{z}-\frac{1}{2}\right)\right]\,.
\end{align}

The infinitely growing sequence of relaxation values, $0<w_1<w_2<...$, are the purely simple, positive roots of the transcendental equation (for an associated heat conduction problem see \cite{Carslawjaeger})
\begin{align} \label{eq:RootsEquation}
	\mu  \sin \left(\frac{w}{2}\right) \cos \left(\frac{w}{2 \mu }\right)+\cos \left(\frac{w}{2}\right) \sin \left(\frac{w}{2 \mu }\right)=0\,,	
\end{align}
associated with the purely imaginary, pairwise conjugate simple poles, $\{\pm i w_k\}$, of the Laplace transformed velocity field $\tilde{v}_x(\overline{z},s)$ constituting a meromorphic function in the complex $s$-plane. The exponentially decaying modes in the series solution reflect the overdamped dynamics described by the momentum diffusion Eqs. (\ref{eq:Stokes}). In the limiting case $\mu=1$ of equal kinematic viscosities, $w_k = \pi k$ with integer values of $k$. For $\mu$ close to or large compared to one, approximate expressions for $w_k(\mu)$ are obtained using perturbation theory. However, these are not useful for the intermediate viscosity ratio $\mu^2 = 4.74$ used in our MPC simulations. For arbitrary $\mu$, we conveniently determined the roots of Eq. (\ref{eq:StationaryFlowField}) using Mathematica.

\begin{figure}[h]
	\centering
	\includegraphics*[width=0.45\textwidth]{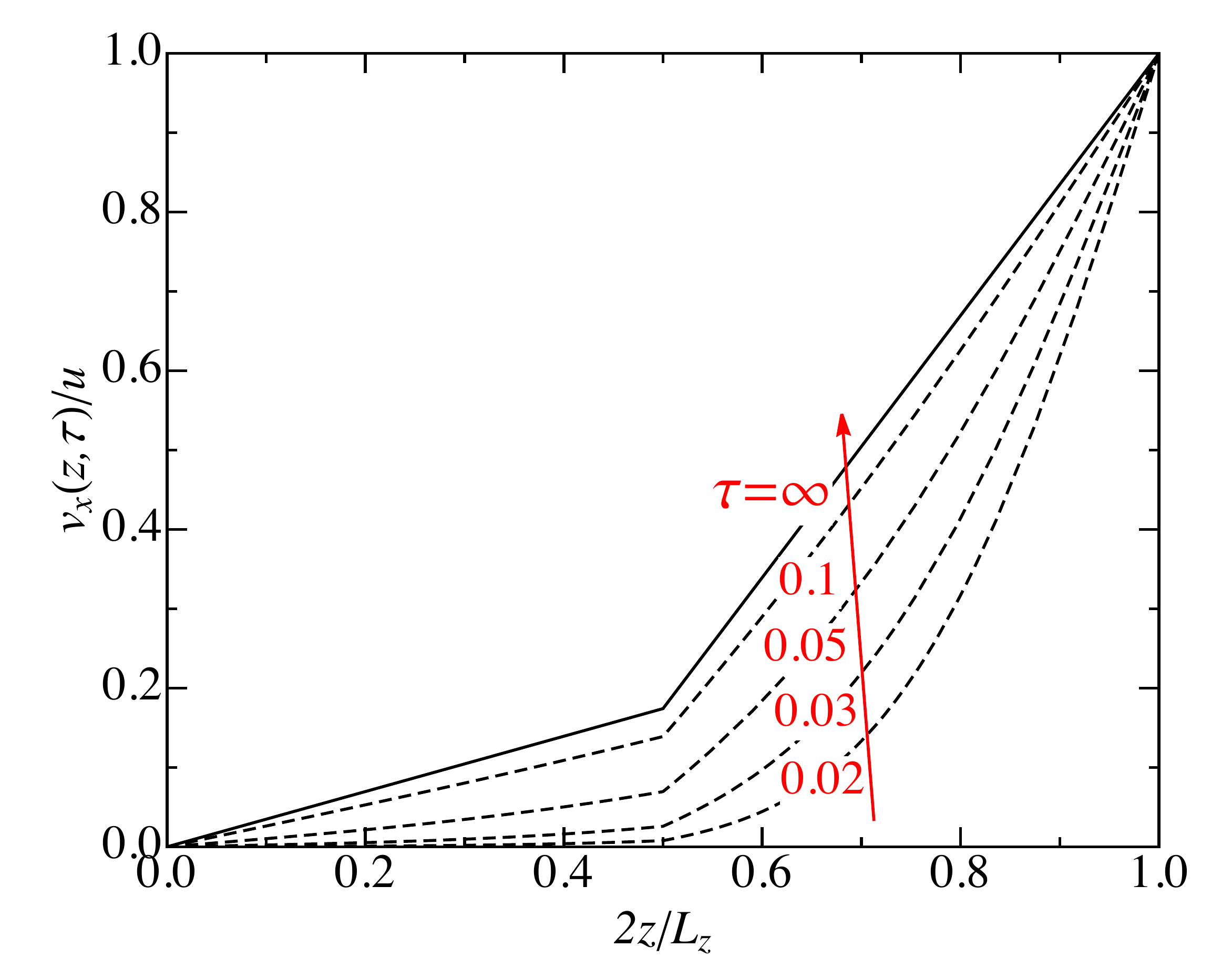}
	\caption{\label{fig:Profiles} Starting flow velocity profiles of the sheared $B$-$A$-$B$ system, obtained from Eqs. (\ref{eq:FlowField}) - (\ref{eq:Denominator}) for the viscosity ratio $\mu^2=4.74$, and for  reduced time values $\tau$ as indicated.}
\end{figure}

Figure \ref{fig:Profiles} depicts the starting flow velocity profiles according to Eqs. (\ref{eq:FlowField}) and (\ref{eq:FlowFieldParts}), for $\mu^2=4.74$ and values of $\tau$ as indicated. For the smallest considered time, $100$ terms in the sum over exponentially decaying modes have been accounted for. The stationary profile is reached  for $\tau > 0.2$.  

The shear stress field follows from the velocity field using $\sigma_{xz}^{A,B}(z,t)=\eta_{A,B}\;\!d u_x^{A,B} (z,t)/dz$. 
The external stress exerted by the walls on the neighboring fluid layer $B$ is obtained in particular as       
\begin{widetext}
  \begin{align}		\label{eq:ExternalStress}
	\frac{\sigma_{xz}^{e}(\tau)}{\sigma_{xz}^{st}}=1+ \frac{2(1+\mu^2)}{\mu}\sum_{k=1}^\infty 
	\frac{\mu \cos(\frac{w_k}{2})\cos(\frac{w_k}{2\;\!\mu})-\sin(\frac{w_k}{2})\sin(\frac{w_k}{2\;\!\mu})}
	{N(w_k)}\times \exp(-w_k^2\;\!\tau )\,,
  \end{align}
\end{widetext}
where
\begin{align} \label{eq:StatWallStress}
	\sigma_{xz}^{st} =  \frac{2 \eta_B u}{L_z}\frac{2\mu^2}{1+\mu^2}
\end{align}
is the spatially uniform stationary long-time stress in the sheared composite system. 
The internal hydrodynamic stress is the spatially averaged shear stress, i.e.,
\begin{align} \label{eq:InternalStress}
	\sigma_{xz}^{i}(t) = \frac{2}{L_z}\int_0^{L_z/4}\!\!dz \sigma^A_{xz}(z,t) 
	+  \frac{2}{L_z}\int_{L_z/4}^{L_z/2}\!\!dz \sigma^B_{xz}(z,t)\,,
\end{align}
which yields
\begin{align} \label{eq:InternalStressResult}\nonumber
	\frac{\sigma^{i}_{xz}(\tau)}{\sigma_{xz}^{st}}
	&=\frac{\mu^2+1}{2\mu^2}\left[1+(\mu^2-1) v_x^{A,B}(\overline{z}=1/2,\tau) \right]\\
	&=1+\frac{2 \left(\mu^4-1\right)}{\mu} \sum_{k=1}^\infty \frac{1}{w_k}  \frac{\sin \left(\frac{w_k}{2 \mu }\right)}{N(w_k)}\exp ( -w_k^2 \tau )\,.
\end{align}
As noted in Sec. 3, in lieu of $\sigma_{xz}^{i,e}(\tau)$, we use the moving time averages 
in the analysis of the  MPC data to smooth out statistical errors (see Eq. (\ref{eq:Cumulative})). The respective analytical stress expressions  follow from Eqs. (\ref{eq:InternalStressResult}) and (\ref{eq:Cumulative})  by the replacement
\begin{align} \label{eq:Replace}
	\exp ( -w_k^2 \tau ) \to \frac{1}{w_k^2 \tau} \left( 1-\exp (-w_k^2 \tau ) \right)\,.
\end{align}

The moving time-averaged stress  converges more slowly toward the stationary stress value $\sigma_{xz}^{st}$ than the stress itself. Both $\sigma_{xz}^{e}(\tau)$ and its moving time average decay monotonically in time toward 
$\sigma_{xz}^{st}$, which reflects the diffusive broadening of the zone of changing fluid velocity with increasing time (see Fig. \ref{fig:Profiles}). At very short times, where the influences of the fluid interfaces and the opposite wall is still negligible, the velocity decays steeply near the wall, from value $u$ at the wall to $0$ slightly off the wall. This implies the approximate  inverse square-root time dependence of the external stress  $\sigma^{e}_{xz}(\tau \ll 1) \approx 2\eta_B u/\left(L_z\sqrt{\pi\nu_B \tau}\;\!\right)$. The inverse square-root short-time divergence is shared by the moving time average stress, except with a twice as large  amplitude for the latter. Different from the external stress, the internal stress is finite at $\tau=0$ where it attains its minimal value  
\begin{align} \label{eq:InternalStressLimit}
	\frac{\sigma^{i}_{xz}(\tau=0)}{\sigma_{xz}^{st}}
	=\frac{\langle\sigma^{i}_{xz}\rangle_{\tau=0}}{\sigma_{xz}^{st}}
	=\frac{1}{2}\left(1+\frac{1}{\mu}^2\right)\,.
\end{align}
For equal viscosities of fluids $A$ and $B$, where the hydrodynamic effect of the $A-B$ fluid interfaces is absent, the intrinsic stress is constant at all times and equal to the stationary stress. 

In the MPC simulation results for the moving time average  stress  depicted in Fig. \ref{fig:shear}(b) for $\mu^2=4.74$, 
time is measured in units of $t_0=a^2/\nu_0$ instead of the reduced time $\tau=t/t_0^\ast$ employed here where $t_0^\ast = (L_z/2)^2/\nu_B$. Using  that $L_z=39 a$ in our MPC study of shear flow, the conversion relation $(t/t_0)\times 10^{-3} \approx 0.44\;\!\tau$ is obtained. We recall furthermore the relation $\sigma_{xz}^{st}= 0.077\;\!\sigma_0$ between stationary stress and MPC thermal stress $\sigma_0=k_BT/a^3$.  
%
\end{document}